\RequirePackage[l2tabu, orthodox]{nag}
\RequirePackage{snapshot}

\documentclass[9pt,onecolumn]{extarticle}

\makeatletter
\if@twocolumn
  \usepackage[dvips,letterpaper,top=0.5in, bottom=0.5in, left=0.75in, right=0.5in,includefoot,heightrounded]{geometry}
\else
  \usepackage[dvips,letterpaper,margin=1in,includefoot,heightrounded]{geometry}
\fi

\usepackage{srcltx}

\usepackage[russian,portuges,english]{babel}

\iflanguage{portuges}
    {\newcommand{\keywordname}{Palavras-chaves}}
    {\newcommand{\keywordname}{Keywords}}

\usepackage{amsmath}
\usepackage{amssymb,amsfonts}

\usepackage{abstract}

\usepackage{graphicx}
\usepackage[usenames,dvipsnames,svgnames,x11names]{xcolor}
\usepackage{subfigure}
\usepackage{psfrag}

\usepackage{booktabs}

\usepackage{setspace}
\usepackage{flushend}

\usepackage{cite}

\usepackage{hyperref}
\usepackage[normalem]{ulem}

\usepackage{enumerate}

\usepackage{multirow}
\usepackage[noend]{algpseudocode}

\usepackage{listings}

\usepackage[short,12hr]{datetime}
       \usepackage{fouriernc}
\makeatletter

\newenvironment{rsmallmatrix}{\null\,\vcenter\bgroup
  \Let@\restore@math@cr\default@tag
  \baselineskip6\ex@ \lineskip1.5\ex@ \lineskiplimit\lineskip
  \ialign\bgroup\hfil$\m@th\scriptstyle##$&&\thickspace\hfil
  $\m@th\scriptstyle##$\crcr
}{%
  \crcr\egroup\egroup\,%
}

\newcommand{\printtitle}{%
\makeatletter
\if@twocolumn

\twocolumn[%
  \maketitle
  \begin{onecolabstract}
    \myabstract
  \end{onecolabstract}
  \begin{center}
    \small
    \textbf{\keywordname}
    \\\medskip
    \mykeywords
  \end{center}
  \bigskip
]
\saythanks
\else
  \maketitle
  \begin{onecolabstract}
    \myabstract
  \end{onecolabstract}
  \begin{center}
    \small
    \textbf{\keywordname}
    \\\medskip
    \mykeywords
  \end{center}
  \bigskip
  \onehalfspacing
\fi
\makeatother
}

\author{%
D.~F.~G.~Coelho%
\thanks{Independent Researcher, Calgary, AB, Canada. Email: (diegofgcoelho@gmail.com)}
\and
R.~J.~Cintra%
\thanks{Signal Processing Group at Statistics,
Universidade Federal de Pernambuco, Recife, PE, Brazil.
Email: (rjdsc@de.ufpe.br)}
\and
F.~M.~Bayer%
\thanks{Departamento de Estat\'{\i}stica and LACESM,
Universidade Federal de Santa Maria, RS, Brazil. Email: (bayer@ufsm.br)}
\and
S.~Kulasekera%
\thanks{Department of Electrical and Computer Engineering,
University of Akron, OH, USA.}
\and
A.~Madanayake%
\thanks{Department of Electrical and Computer Engineering, Florida International University, FL, USA. Email: (amadanay@fiu.edu)}
\and
P.~A.~C.~Martinez%
\thanks{Signal Processing Group at Statistics,
Universidade Federal de Pernambuco, Recife, Brazil
and
Friedrich-Alexander-Universit\"at Erlangen-N\"urnberg, Erlangen,
Germany.}
\and
T.~L.~T.~Silveira%
\thanks{Programa de P\'os-Gradua\c c\~ao em Computa\c c\~ao,
Universidade Federal do Rio Grande do Sul (UFRGS), Porto~Alegre, RS,
Brazil. Email: (thiago@inf.ufsm.br)}
\and
R.~S.~Oliveira%
\thanks{Programa de P\'os-Gradua\c{c}\~ao em Engenharia El\'etrica, Universidade Federal de Pernambuco, Recife, Brazil.}
\and
V.~S.~Dimitrov%
\thanks{Department of Electrical and Computer Engineering,
University of Calgary, Calgary, AB, Canada. Email: (vdvsd103@gmail.com)}
}

\title{%
Low-Complexity Loeffler DCT Approximations
for
Image and Video Coding}

\newcommand{\myabstract}{%
This paper
introduced
a matrix parametrization method  based
on the Loeffler discrete cosine transform (DCT) algorithm.
As a result,
a new class of eight-point DCT approximations
was proposed,
capable of
unifying the mathematical formalism
of several eight-point DCT approximations archived in the literature.
Pareto-efficient DCT approximations
are obtained through multicriteria optimization,
where
computational complexity, proximity, and coding performance
are considered.
Efficient approximations
and their scaled 16- and 32-point versions
are embedded into
image and video encoders,
including a JPEG-like codec
and
\mbox{H.264}/AVC and \mbox{H.265}/HEVC
standards.
Results are compared to the unmodified standard codecs.
Efficient approximations
are mapped and implemented on a Xilinx VLX240T FPGA
and evaluated for area, speed, and power consumption.
}

\newcommand{\mykeywords}{%
Discrete cosine transform; approximation; multicriteria optimization; image/video compression
}

\date{}

\begin{document}

\printtitle

\section{Introduction}

Discrete time transforms
have a major role in signal-processing theory
and application.
In particular,
tools
such as the discrete Haar,
Hadamard,
and discrete Fourier transforms,
and
several discrete trigonometrical
transforms~\cite{Ahmed1975,Britanak2007}
have contributed
to various image-processing techniques~\cite{Kolaczyk1997,Qiang2011,Bouguezel2013,Martucci1993}.
Among such transformations,
the discrete cosine transform (DCT)
of type~II
is widely regarded as a pivotal tool
for image compression, coding, and analysis~\cite{Oppenheim1999,Gonzalez2001,Britanak2007}.
This is because the DCT
closely
approximates
the Karhunen--Lo\`eve transform (KLT)
which can optimally decorrelate
highly correlated stationary Markov-I
signals~\cite{Britanak2007}.

Indeed, the
recent literature
reveals a significant number of works linked to DCT computation.
Some noteworthy topics are:
(i)~cosine--sine decomposition to compute the eight-point DCT~\cite{Parfieniuk2015};
(ii)~low-complexity-pruned eight-point DCT approximations for image encoding~\cite{A.Coutinho2015};
(iii)~improved eight-point approximate DCT for image and video compression requiring only 14 additions~\cite{Potluri2014};
(iv)~HEVC multisize DCT hardware with constant throughput, supporting heterogeneous coding unities~\cite{Goebel2016};
(v)~approximation of feature pyramids in the DCT domain and its application to pedestrian detection~\cite{Naiel2016};
(vi)~performance analysis of DCT and discrete wavelet transform (DWT) audio watermarking based on singular value decomposition~\cite{Lalitha2016};
(vii)~adaptive approximated DCT architectures for HEVC~\cite{Masera2016};
(viii)~improved Canny edge detection algorithm based on DCT~\cite{Zhao2015}; and
(ix) DCT-inspired feature transform for image retrieval and reconstruction~\cite{Wang2016}.
In fact,
several current image- and video-coding schemes
are based on the DCT~\cite{Bhaskaran1995},
such as
 JPEG~\cite{Wallace1992},
MPEG-1~\cite{Roma2007},
\mbox{H.264}~\cite{Wiegand2003},
and
HEVC~\cite{Pourazad2012}.
In particular,
the \mbox{H.264} and HEVC codecs
employ
low-complexity discrete transforms based
on the eight-point DCT.
The eight-point DCT has also been applied to dedicated image-compression systems implemented in large web servers with promising results~\cite{Horn2016}.
As a consequence,
several algorithms for the eight-point DCT
have been proposed,
such as:
 Lee DCT factorization~\cite{Lee1984},
Arai DCT scheme~\cite{Arai1988},
Feig--Winograd algorithm~\cite{Feig1992},
and the Loeffler DCT algorithm~\cite{Loeffler1989}.
Among these methods,
the Loeffler DCT algorithm~\cite{Loeffler1989}
has the distinction of achieving
the theoretical lower bound for
 DCT multiplicative complexity~\cite{Duhamel1987,Heideman1988}.

Because the computational complexity lower bounds of the DCT
have been achieved~\cite{Duhamel1987},
the research community resorted to approximation
techniques to further reduce the cost of DCT calculation.
Although not capable of providing exact computation,
approximate transforms can furnish
very close computational results
at significantly smaller computational cost.
Early approximations for the DCT
were introduced by Haweel~\cite{Haweel2001}.
Since then, several DCT approximations
have been proposed.
In Reference~\cite{Lengwehasatit2004},
Lengwehasatit and Ortega
introduced a scalable approximate DCT
that can be regarded as a benchmark approximation~\cite{Lengwehasatit2004, Cintra2011, Cintra2012, Bouguezel2008, Bouguezel2008a, Bouguezel2010, Bouguezel2013, Bouguezel2011, Cintra2011a, Cintra2014, Zhang2007}.
Aiming at image coding for data compression,
a~series of approximations have been proposed
by Bouguezel-Ahmad-Swamy (BAS)~\cite{Bouguezel2008, Bouguezel2008a, Bouguezel2009, Bouguezel2010, Bouguezel2011, Bouguezel2013}.
Such approximations offer very low complexity and
good coding performance~\cite{Cintra2014, Britanak2007}.

The methods for deriving DCT approximation
include:
(i)~application of simple functions, such
as signum, rounding-off, truncation, ceil, and floor, to approximate
the elements of the exact DCT matrix~\cite{Haweel2001,Cintra2014};
(ii)~scaling and rounding-off~\cite{Britanak2007,Cintra2011, Chen2002, JieLiang2000, JieLiang2001, Cintra2014,Malvar1986, Malvar1987, Malvar1987a};
(iii)~brute-force computation over
reduced search space~\cite{Cintra2014,Cintra2011a};
(iv)~inspection~\cite{Cintra2012, Bouguezel2008, Bouguezel2008a, Bouguezel2009};
(v)~single-variable matrix parametrization of existing approximations~\cite{Bouguezel2011};
(vi)~pruning techniques~\cite{Kouadria2013};
and
(vii)~derivations based on other low-complexity matrices~\cite{Bouguezel2013}.
The above-mentioned methods are capable of supplying
single or very few approximations.
In fact,
a systematized approach for obtaining
a large number
of matrix approximations
and
a unifying scheme is lacking.

The goal of this paper is two-fold.
First we aim at
unifying
the matrix formalism
of several eight-point DCT approximations archived in the literature.
For that, we consider the eight-point Loeffler algorithm as a general structure equipped with a parametrization
of the multiplicands.
This approach allows the definition
of a matrix subspace where
a large number of approximations could be derived.
Second,
we propose an optimization problem
over the introduced matrix subspace
in order to discriminate the best approximations
according to several well-known figures of merit.
This discrimination is important from the application point of view.
It
allows
the user
to select the transform that fits best
to their application in
terms of balancing performance and complexity.
The optimally found approximations
are subject to mathematical
assessment
and embedding
into
image- and video-encoding schemes,
including
the
\mbox{H.264}/AVC
and
the
\mbox{H.265}/HEVC
standards.
Third,
we introduce hardware architecture based on
optimally found approximations
realized in
field programmable gate array~(FPGA).
Although there are several subsystems in a video and image codec,
this work is solely concentrated  on the discrete transform subsystem.

The paper unfolds as follows.
Section~\ref{DCTappr}
introduces
a novel DCT parametrization
based on the Loeffler DCT algorithm.
We provide the mathematical background
and
matrix properties,
such as invertibility,
orthogonality,
and
orthonormalization, are examined.
Section~\ref{Loefflersearch}
reviews the criteria employed
for identifying and assessing DCT approximations,
such as
proximity and coding measures,
and
computational complexity.
In Section~\ref{multicriteriaoptimization},
we propose
a multicriteria-optimization
problem aiming at deriving optimal approximation
subject to Pareto efficiency.
Obtained transforms are sought to be
comprehensively  assessed
and compared with state-of-the-art competitors.
Section~\ref{sec:experiments}
reports the results of embedding
the obtained transforms
into a JPEG-like encoder,
as well as
in \mbox{H.264}/AVC and \mbox{H.265}/HEVC
video standards.
In Section~\ref{sec:fpga},
an FPGA hardware implementation
of the optimal transformations
is detailed, and the
usual FPGA implementation metrics are reported.
Section~\ref{conclusion} presents our final remarks.

\section{DCT Parametrization and Matrix Space}
\label{DCTappr}
\vspace{-6PT}

\subsection{DCT Matrix Factorization}

The  type II DCT
is defined according
to the following
linear transformation matrix~\cite{Britanak2007, Gonzalez2001}:
\begin{align}
\mathbf{C}_\text{DCT}
&
=
\frac{1}{2} \left[
\begin{rsmallmatrix}
c_4 & c_4  &     c_4     &  c_4  &     c_4   &    c_4    &   c_4  &     c_4  \\
     c_1   &   c_3  &     c_5  &     c_7    &   -c_7&      -c_5&      -c_3&      -c_1 \\
     c_2   &   c_6   &    -c_6 &     -c_2   &   -c_2&      -c_6  &    c_6  &     c_2  \\
     c_3   &   -c_7  &    -c_1 &     -c_5   &   c_5 &      c_1   &    c_7  &     -c_3 \\
     c_4  &    -c_4  &    -c_4 &     c_4    &   c_4 &      -c_4  &    -c_4 &     c_4  \\
     c_5   &   -c_1  &    c_7  &     c_3    &   -c_3&      -c_7  &    c_1  &     -c_5 \\
     c_6   &   -c_2  &    c_2  &     -c_6  &    -c_6&      c_2   &    -c_2 &     c_6  \\
     c_7   &   -c_5  &    c_3  &     -c_1   &   c_1 &      -c_3   &   c_5  &     -c_7\\
\end{rsmallmatrix}
\right]
,
\end{align}
where~$c_k = \cos(k\pi/16)$,
$k=1,2,\ldots,7$.
Because several entries of~$\mathbf{C}_\text{DCT}$
are not rational,
they
are often truncated/rounded and
represented in floating-point arithmetic~\cite{Oppenheim1999, Manassah2001},
which requires
demanding computational costs
when compared with fixed-point
schemes~\cite{Blahut2010, Oppenheim1999, Cintra2012, Bouguezel2013}.

Fast algorithms
can minimize the number of arithmetic operations
required for the DCT computation~\cite{Britanak2007,Oppenheim1999}.
A number of fast algorithms have been
proposed for the~eight-point
DCT~\cite{Lee1984, Arai1988, Feig1992}.
The vast majority of DCT algorithms consist of
the following factorization~\cite{Britanak2007}:
\begin{align}
\mathbf{C}_\text{DCT} & = \mathbf{P}\cdot \mathbf{M} \cdot \mathbf{A}
,
\end{align}
where
$\mathbf{A}$ is the additive matrix that
represents a set of butterfly operations,
$\mathbf{M}$ is a multiplicative matrix,
and
$\mathbf{P}$ is a permutation matrix
that simply
rearranges the output components
to natural order.
Matrix~$\mathbf{A}$ is often fixed~\cite{Britanak2007} and
given by:
\begin{align}
\mathbf{A} & = \left[
\begin{rsmallmatrix}
1&\phantom{-}0&\phantom{-}0&\phantom{-}0&\phantom{-}0&\phantom{-}0&\phantom{-}0&\phantom{-}1\\
0&1&0&0&0&0&1&0\\
0&0&1&0&0&1&0&0\\
0&0&0&1&1&0&0&0\\
0&0&0&1&-1&0&0&0\\
0&0&1&0&0&-1&0&0\\
0&1&0&0&0&0&-1&0\\
1&0&0&0&0&0&0&-1\\
\end{rsmallmatrix}
\right].
\end{align}
Multiplicative matrix~$\mathbf{M}$
can be further factorized.
Since matrix factorization is not unique,
each fast algorithm is
linked to
a particular factorization of~$\mathbf{M}$.
Finally,
the permutation matrix is given~below:
\begin{align}
\mathbf{P} & =
\left[
\begin{rsmallmatrix}
1&0&0&0&0&0&0&0\\
0&0&0&0&0&0&0&1\\
0&0&1&0&0&0&0&0\\
0&0&0&0&0&1&0&0\\
0&1&0&0&0&0&0&0\\
0&0&0&0&0&0&1&0\\
0&0&0&1&0&0&0&0\\
0&0&0&0&1&0&0&0\\
\end{rsmallmatrix}
\right]
.
\end{align}

Among the DCT fast algorithms,
the
Loeffler DCT
achieves the theoretical lower bound of
the multiplicative complexity for~$8$-point DCT,
which consists of 11~multiplications~\cite{Loeffler1989}.
In this work, multiplications by irrational quantities
are sought to be substituted with trivial multipliers,
representable by simple bit-shifting operations
(Sections~\ref{subsec:para} and~\ref{subsec:computational_cost}).
Thus, we expect that approximations based
on Loeffler DCT could
generate low-complexity approximations.
Therefore, the Loeffler DCT algorithm
was separated
as the starting point to devise
new DCT approximations.
The Loeffler DCT employs a scaled DCT
with the following transformation matrix:
\begin{align}
\mathbf{C}_\text{Loeffler-DCT}
=
2\sqrt{2}
\cdot
\mathbf{C}_\text{DCT}
.
\end{align}
Such scaling eliminates one multiplicand
because
$2\sqrt{2} \cdot c_4 = 1$.
Therefore,
we can write the following~expression:
\begin{align}
\label{eq:dct-factorization}
\mathbf{C}_\text{Loeffler-DCT}
=
\mathbf{P}
\cdot
\mathbf{M}'
\cdot
\mathbf{A}
,
\end{align}
where
\begin{align}
\label{equation-multiplicative-matrix}
\mathbf{M}'
=
&
2\sqrt{2}\cdot\mathbf{M}\nonumber
\\
=
&
\left[
\begin{rsmallmatrix}
1&1&1&1&0&0&0&0\\
1&-1&-1&1&0&0&0&0\\
\sqrt{2}c_2&\sqrt{2}c_6&-\sqrt{2}c_6&-\sqrt{2}c_2&0&0&0&0\\
\sqrt{2}c_6&-\sqrt{2}c_2&\sqrt{2}c_2&-\sqrt{2}c_6&0&0&0&0\\
0&0&0&0&-\sqrt{2}c_1&\sqrt{2}c_3&-\sqrt{2}c_5&\sqrt{2}c_7\\
0&0&0&0&-\sqrt{2}c_5&-\sqrt{2}c_1&-\sqrt{2}c_7&\sqrt{2}c_3\\
0&0&0&0&\sqrt{2}c_3&\sqrt{2}c_7&-\sqrt{2}c_1&\sqrt{2}c_5\\
0&0&0&0&\sqrt{2}c_7&\sqrt{2}c_5&\sqrt{2}c_3&\sqrt{2}c_1\\
\end{rsmallmatrix}
\right]
.
\end{align}

Matrix~$\mathbf{M}'$ carries all multiplications by irrational quantities required by Loeffler fast algorithm.
It can be further decomposed as:
\begin{align}
\label{equation-multiplicative-matrix}
\mathbf{M}'
=
&
\mathbf{B}\cdot \mathbf{C}\cdot \mathbf{D},
\end{align}
where
\begin{align}
\mathbf{B}
=
&
\left[
\begin{rsmallmatrix}
1&\phantom{-}0&0&1&0&0&\phantom{-}0&\phantom{-}0\\
0&1&1&0&0&0&0&0\\
0&1&-1&0&0&0&0&0\\
1&0&0&-1&0&0&0&0\\
0&0&0&0&c_3&0&0&c_5\\
0&0&0&0&0&c_1&c_7&0\\
0&0&0&0&0&-c_7&c_1&0\\
0&0&0&0&-c_5&0&0&c_3\\
\end{rsmallmatrix}
\right]
,
\end{align}
\begin{align}
\mathbf{C}
=
&
\left[
\begin{rsmallmatrix}
1&1&0&0&0&0&0&0\\
1&-1&0&0&0&0&0&0\\
0&0&\sqrt{2}c_6&\sqrt{2}c_2&\phantom{-}0&\phantom{-}0&\phantom{-}0&\phantom{-}0\\
0&0&-\sqrt{2}c_2&\sqrt{2}c_6&0&0&0&0\\
0&0&0&0&1&0&1&0\\
0&0&0&0&0&-1&0&1\\
0&0&0&0&1&0&-1&0\\
0&0&0&0&0&1&0&1\\
\end{rsmallmatrix}
\right]
,
\end{align}
and
\begin{align}
\mathbf{D}
=
&
\left[
\begin{rsmallmatrix}
1&\phantom{-}0&\phantom{-}0&\phantom{-}0&\phantom{-}0&\phantom{-}0&\phantom{-}0&\phantom{-}0\\
0&1&0&0&0&0&0&0\\
0&0&1&0&0&0&0&0\\
0&0&0&1&0&0&0&0\\
0&0&0&0&-1&0&0&1\\
0&0&0&0&0&\sqrt{2}&0&0\\
0&0&0&0&1&0&\sqrt{2}&0\\
0&0&0&0&1&0&0&1\\
\end{rsmallmatrix}
\right]
.
\end{align}
In order to achieve the minimum multiplicative complexity, the Loeffler fast algorithm uses fast rotation for the rotation blocks in matrix~$\mathbf{B}$ and~$\mathbf{C}$~\cite{Blahut2010, Britanak2007}.
Since each of the three fast rotations requires three multiplications, and we have the two additional multiplications on matrix~$\mathbf{D}$, the Loeffler fast algorithm requires a total of~$11$ multiplications.

\subsection{Loeffler DCT Parametrization}
\label{subsec:para}

 DCT factorization
suggests matrix parametrization.
In fact,
replacing multiplicands
$\sqrt{2}\cdot c_i$,
$i\in\{1,2,3,5,6,7\}$
in Matrix~\eqref{equation-multiplicative-matrix}
by
parameters~$\alpha_k$,
$k=1,2,\ldots,6$,
respectively,
yields
the following
parametric matrix:
\begin{align}
\mathbf{M}_{\pmb{\alpha}}
& =
\left[
\begin{rsmallmatrix}
1&1&1&1&0&0&0&0\\
1&-1&-1&1&0&0&0&0\\
\alpha_2&\alpha_5&-\alpha_5&-\alpha_2&0&0&0&0\\
\alpha_5&-\alpha_2&\alpha_2&-\alpha_5&0&0&0&0\\
0&0&0&0&-\alpha_1&\alpha_3&-\alpha_4&\alpha_6\\
0&0&0&0&-\alpha_4&-\alpha_1&-\alpha_6&\alpha_3\\
0&0&0&0&\alpha_3&\alpha_6&-\alpha_1&\alpha_4\\
0&0&0&0&\alpha_6&\alpha_4&\alpha_3&\alpha_1\\
\end{rsmallmatrix}
\right]
,
\end{align}
where subscript~$\pmb{\alpha}=\begin{bmatrix}\alpha_1 & \alpha_2 & \cdots & \alpha_6 \end{bmatrix}^\top$
denotes a real-valued parameter vector.
Mathematically,
the following
mapping is introduced:
\begin{equation}
\label{matrixmapping}
\begin{split}
f: \mathbb{R}^6 & \longrightarrow \mathcal{M}(8) \\
\pmb{\alpha} & \longmapsto
\mathbf{T}_{\pmb{\alpha}}
=
\mathbf{P}\cdot \mathbf{M}_{\pmb{\alpha}} \cdot \mathbf{A},
\end{split}
\end{equation}
where
$\mathcal{M}(8)$
represents the space of $8 \times 8$ matrices
over the real numbers~\cite{Halmos2013}.
Mapping $f(\cdot)$
results
in image set~$\mathcal{C}(8) \subset \mathcal{M}(8)$~\cite{Halmos2013}
that contains 8$\times$8 matrices with the DCT matrix symmetry.
In particular,
for
$\pmb{\alpha}_0 = \sqrt{2} \cdot \begin{bmatrix}c_1 & c_2 & c_3 & c_5 & c_6 & c_7\end{bmatrix}^\top$,
we have that $f(\pmb{\alpha}_0) = \mathbf{C}_\text{Loeffler-DCT}$.
Other examples
are
$\pmb{\alpha}_1=\begin{bmatrix}1&1&1&1&1&1&1\end{bmatrix}^\top$
and
\mbox{$\pmb{\alpha}_2=1/2\cdot \begin{bmatrix}1&2&1&1&1&1&2\end{bmatrix}^\top$},
which result in the following matrices:
\begin{align}
\mathbf{T}_{\pmb{\alpha}_1}
=
\left[
\begin{rsmallmatrix}
1 &1 &1 &1 &1 &1 &1 &1 \\
1 &1 &1 &1 &-1 &-1 &-1 &-1 \\
1 &1 &-1 &-1 &-1 &-1 &1 &1 \\
1 &-1 &-1 &-1 &1 &1 &1 &-1 \\
1 &-1 &-1 &1 &1 &-1 &-1 &1 \\
1 &-1 &1 &1 &-1 &-1 &1 &-1 \\
1 &-1 &1 &-1 &-1 &1 &-1 &1 \\
1 &-1 &1 &-1 &1 &-1 &1 &-1
\end{rsmallmatrix}
\right]
\quad
\text{and}
\quad
\mathbf{T}_{\pmb{\alpha}_2}
=
\frac{1}{2}
\left[
\begin{rsmallmatrix}
2 &2 &2 &2 &2 &2 &2 &2 \\
1 &1 &1 &2 &-2 &-1 &-1 &-1 \\
2 &1 &-1 &-2 &-2 &-1 &1 &2 \\
1 &-2 &-1 &-1 &1 &1 &2 &-1 \\
2 &-2 &-2 &2 &2 &-2 &-2 &2 \\
1 &-1 &2 &1 &-1 &-2 &1 &-1 \\
1 &-2 &2 &-1 &-1 &2 &-2 &1 \\
2 &-1 &1 &-1 &1 &-1 &1 &-2
\end{rsmallmatrix}
\right]
,
\end{align}
respectively.
Although the above matrices
have low complexity,
they may not necessarily lead to a good transform matrix
in terms of mathematical properties and coding capability.

Hereafter,
we adopt the following notation:
\begin{align}
\mathbf{M}_{\pmb{\alpha}} & =
\left[
\begin{rsmallmatrix}
\mathbf{E}_{\pmb{\alpha}} & \mathbf{0}_{4} \\
\mathbf{0}_{4} & \mathbf{O}_{\pmb{\alpha}} \\
\end{rsmallmatrix}
\right],
\end{align}
where
$\mathbf{0}_4$ is the the $4 \times 4$ null matrix,
and
\begin{align}
\mathbf{E}_{\pmb{\alpha}} & =
\left[
\begin{rsmallmatrix}
1&1&1&1\\
1&-1&-1&1\\
\alpha_2&\alpha_5&-\alpha_5&-\alpha_2\\
\alpha_5&-\alpha_2&\alpha_2&-\alpha_5\\
\end{rsmallmatrix}
\right]
\:\:\:
\text{and}
\:\:\:
\mathbf{O}_{\pmb{\alpha}} =
\left[
\begin{rsmallmatrix}
-\alpha_1&\alpha_3&-\alpha_4&\alpha_6\\
-\alpha_4&-\alpha_1&-\alpha_6&\alpha_3\\
\alpha_3&\alpha_6&-\alpha_1&\alpha_4\\
\alpha_6&\alpha_4&\alpha_3&\alpha_1\\
\end{rsmallmatrix}
\right].
\end{align}
Submatrices~$\mathbf{E}_{\pmb{\alpha}}$ and $\mathbf{O}_{\pmb{\alpha}}$
compute the even and odd index DCT components,
respectively.

\subsection{Matrix Inversion}

The
inverse of $\mathbf{T}_\alpha$
is directly given by:
\begin{align}
\mathbf{T}_\alpha^{-1} = \mathbf{A}^{-1} \cdot \mathbf{M}_{\pmb{\alpha}}^{-1}\cdot  \mathbf{P}^{-1}.
\end{align}
Because
$\mathbf{A}^{-1}  = \frac{1}{2} \mathbf{A}^{\top} = \frac{1}{2}\mathbf{A}$
and
$\mathbf{P}^{-1}  = \mathbf{P}^{\top}$
are well-defined, nonsingular matrices,
we need only check the invertibility
of
$\mathbf{M}_{\pmb{\alpha}}$~\cite{Graham1989}.
By means of symbolic computation,
we obtain:
\begin{align}
\mathbf{M}_{\pmb{\alpha}}^{-1}
=
\frac{1}{4}
\cdot
(\mathbf{M}_{\pmb{\alpha'}})^{\top}
,
\label{eq:invM}
\end{align}
where
$\pmb{\alpha}' =
\begin{bmatrix}
\alpha'_1 & \alpha'_2 & \alpha'_3 & \alpha'_4 & \alpha'_5 & \alpha'_6
\end{bmatrix}^\top$
with coefficients equals to:
\begin{equation}
\small
\begin{split}
\alpha'_1 & = -4( \alpha_1^3 + 2 \alpha_1 \alpha_3 \alpha_4 + \alpha_1 \alpha_6^2 + \alpha_3^2 \alpha_6 - \alpha_4^2 \alpha_6)
/
\operatorname{det}(\mathbf{O}_{\pmb{\alpha}})
,
\\
\alpha'_2 & =  -16\alpha_2
/
\operatorname{det}(\mathbf{E}_{\pmb{\alpha}})
,
\\
\alpha'_3 &
=
 4(-\alpha_1^2 \alpha_4 - 2 \alpha_1 \alpha_3 \alpha_6 - \alpha_3^3 - \alpha_3 \alpha_4^2 + \alpha_4 \alpha_6^2)
/
\operatorname{det}(\mathbf{O}_{\pmb{\alpha}})
,
\\
\alpha'_4
& =
-4(\alpha_1^2 \alpha_3 - 2 \alpha_1 \alpha_4 \alpha_6 + \alpha_3^2 \alpha_4 - \alpha_3 \alpha_6^2 + \alpha_4^3)
/
\operatorname{det}(\mathbf{O}_{\pmb{\alpha}})
,
\\
\alpha'_5 & = -16\alpha_5
/
\operatorname{det}(\mathbf{E}_{\pmb{\alpha}})
,
\\
\alpha'_6 & = 4(- \alpha_1^2 \alpha_6 - \alpha_1 \alpha_3^2 + \alpha_1 \alpha_4^2 + 2
\alpha_3 \alpha_4 \alpha_6 - \alpha_6^3)
/
\operatorname{det}(\mathbf{O}_{\pmb{\alpha}})
,
\end{split}
\end{equation}
where
$\operatorname{det}(\cdot)$ returns the determinant.

Note that the expression in Matrix~\eqref{eq:invM} implies that the inverse of the matrix~$\mathbf{M}_{\pmb{\alpha}}$ is a matrix with the same structure, whose coefficients are a function of the parameter vector~$\pmb{\alpha}$.
For the matrix inversion
to be well-defined,
we must have:
$\operatorname{det}(\mathbf{E}_{\pmb{\alpha}}) \cdot \operatorname{det}(\mathbf{O}_{\pmb{\alpha}}) \neq 0
$.
By explicitly computing
$\operatorname{det}(\mathbf{E}_{\pmb{\alpha}})$
and
$\operatorname{det}(\mathbf{O}_{\pmb{\alpha}})$,
we obtain the following condition for matrix inversion:
\begin{align}
\label{detU}
\begin{cases}
\alpha_2^2+\alpha_5^2
\neq
0
,
\\
\frac
{(\alpha_1^2+\alpha_6^2)^2-4\alpha_3\alpha_4(\alpha_6^2-\alpha_1^2)}
{(\alpha_3^2+\alpha_4^2)^2-4\alpha_1\alpha_6(\alpha_4^2-\alpha_3^2)}
\neq
-1
.
\end{cases}
\end{align}

\subsection{Orthogonality}

In this paper,
we adopt the following definitions.
A matrix~$\mathbf{T}$ is orthonormal if
$\mathbf{T} \cdot \mathbf{T}^\top$
is an identity matrix~\cite{Shores2007}.
If product~$\mathbf{T} \cdot \mathbf{T}^\top$
is a diagonal matrix,
$\mathbf{T}$ is said to be orthogonal.
For $\mathbf{T}_{\pmb{\alpha}}$,
we have that symbolic computation yields:
\begin{align}
\mathbf{T}_{\pmb{\alpha}}\cdot \mathbf{T}_{\pmb{\alpha}}^{\top} & =
\left[
\begin{rsmallmatrix}
8&0&0&0&0&0&0&0\\
0&2s_1&0&-2d&0&2d&0&0\\
0&0&2s_0&0&0&0&0&0\\
0&-2d&0&2s_1&0&0&0&2d\\
0&0&0&0&8&0&0&0\\
0&2d&0&0&0&2s_1&0&2d\\
0&0&0&0&0&0&2s_0&0\\
0&0&0&2d&0&2d&0&2s_1\\
\end{rsmallmatrix}
\right],
\end{align}
where
$s_0  = 2(\alpha_2^2+\alpha_5^2)$,
$s_1  = \alpha_1^2+\alpha_3^2+\alpha_4^2+\alpha_6^2$,
and
$d  = \alpha_1(\alpha_4-\alpha_3)+\alpha_6(\alpha_4+\alpha_3)$.
Thus,
if~$d=0$,
then
the transform~$\mathbf{T}_{\pmb{\alpha}}$ is orthogonal.

\subsection{Near Orthogonality}

Some important and well-known DCT approximations are
nonorthogonal~\cite{Bouguezel2008, Haweel2001}.
Nevertheless,
such transformations are nearly orthogonal~\cite{Tablada2015,Cintra2014}.
Let $\mathbf{A}$ be a square matrix.
Deviation from orthogonality
can be quantified
according to
the deviation from diagonality~\cite{Cintra2014}
of
$\mathbf{A}\cdot\mathbf{A}^\top$,
which is
given by
the following expression:
\begin{align}
\delta(\mathbf{A}\cdot \mathbf{A}^{\top})
& =
1-
\frac
{
\|\operatorname{diag}(\mathbf{A}\cdot \mathbf{A}^{\top})\|_{\mathsf{F}}^2
}
{
\|
\mathbf{A}\cdot \mathbf{A}^{\top}
\|_{\mathsf{F}}^2}
,
\end{align}
where
$\operatorname{diag}(\cdot)$
returns a diagonal matrix with the diagonal elements of its argument
and
$\|\cdot\|_{\mathsf{F}}$ denotes the Frobenius norm~\cite{Britanak2007}.
Therefore,
considering~$\mathbf{T}_{\pmb{\alpha}}$,
we obtain:
\begin{align}
\delta(\mathbf{T}_{\pmb{\alpha}}\cdot \mathbf{T}_{\pmb{\alpha}}^{\top})
& =
1
-
\frac{1}
{
1
+
\frac
{32 d^2}
{128+8s_0^2+16s_1^2}
}
.
\end{align}

Nonorthogonal transforms
have been recognized as
useful tools.
The signed DCT (SDCT) is a particularly
relevant DCT approximation~\cite{Haweel2001}
and
its
deviation from orthogonality is $0.20$.
We adopt such deviation
as a reference value to discriminate
nearly orthogonal matrices.
Thus,
for
$\mathbf{T}_{\pmb{\alpha}}$,
we obtain the following
criterion for near orthogonality:
\begin{align}
\label{equation-nearly-orthogonality-criterion}
0<d^2 & \leq
1
+
\frac{s_0^2}{16}
+
\frac{s_1^2}{8}
.
\end{align}

\subsection{Orthonormalization}

Discrete transform approximations  are often sought
to be orthonormal.
Orthogonal transformations
can be orthonormalized
as described in References~\cite{Shores2007,Britanak2007,Cintra2012}.
Based on
 polar decomposition~\cite{Strang2005},
 orthonormal or nearly orthonormal
matrix~$\hat{\mathbf{C}}_{\pmb{\alpha}}$
linked
to
$\mathbf{T}_{\pmb{\alpha}}$
is furnished~by:
\begin{align}
\hat{\mathbf{C}}_{\pmb{\alpha}}
=
\begin{cases}
\mathbf{S}_{\pmb{\alpha}} \cdot \mathbf{T}_{\pmb{\alpha}},
&
\text{if $d=0$,}
\\
\tilde{\mathbf{S}}_{\pmb{\alpha}} \cdot \mathbf{T}_{\pmb{\alpha}},
&
\text{if \eqref{equation-nearly-orthogonality-criterion} holds true.}
\end{cases}
\end{align}
where
$\mathbf{S}_{\pmb{\alpha}} = \sqrt[]{\left(\mathbf{T}_{\pmb{\alpha}}\cdot \mathbf{T}_{\pmb{\alpha}}^{\top}\right)^{-1}}$,
$\tilde{\mathbf{S}}_{\pmb{\alpha}}  = \sqrt[]{\operatorname{diag}\left(\mathbf{T}_{\pmb{\alpha}}\cdot \mathbf{T}_{\pmb{\alpha}}^{\top}\right)^{-1}}$,
and
$\sqrt{\cdot}$
is the matrix square root~\cite{Britanak2007}.

\section{Assessment Criteria}
\label{Loefflersearch}

In this section,
we describe
the selected figures of merit
for assessing
the
performance and complexity
of a given DCT approximation.
We separated the following performance metrics:
(i)~total error energy~\cite{Cintra2011, Tablada2015};
(ii)~mean square error (MSE)~\cite{Britanak2007,Manassah2001};
(iii)~unified coding gain~\cite{Britanak2007,Tablada2015},
and
(iv)~transform efficiency~\cite{Britanak2007}.
For
computational complexity
assessment,
we adopted arithmetic operation counts
as figures of merit.

\subsection{Performance Metrics}
\vspace{-6pt}

\subsubsection{Total Error Energy}

Total error energy quantifies
the error between matrices
in a Euclidean distance way.
This measure is given by References~\cite{Cintra2011, Tablada2015}:
\begin{align}
\epsilon
\left(
\hat{\mathbf{C}}_{\pmb{\alpha}}
\right)
&
=
\pi \cdot \|
\mathbf{C}_\text{Loeffler-DCT}
-
\hat{\mathbf{C}}_{\pmb{\alpha}}
\|_{\mathsf{F}}^2
.
\end{align}

\subsubsection{Mean Square Error}

The MSE
of given matrix approximation~$\hat{\mathbf{C}}_{\pmb{\alpha}}$
is furnished by:
\begin{equation}
\begin{split}
\operatorname{MSE}
\left(
\hat{\mathbf{C}}_{\pmb{\alpha}}
\right)
=
\frac{1}{8}
&
\cdot
\operatorname{trace}
\left(
(
\mathbf{C}_\text{Loeffler-DCT}
-
\hat{\mathbf{C}}_{\pmb{\alpha}}
)
\cdot
\mathbf{R}_{\mathbf{xx}}
\right.
\\
&
\cdot
\left.
(
\mathbf{C}_\text{Loeffler-DCT}
-
\hat{\mathbf{C}}_{\pmb{\alpha}}
)^\top
\right)
,
\end{split}
\end{equation}
where~$\mathbf{R}_{\mathbf{xx}}$ represents the autocorrelation matrix
of a Markov I stationary process with correlation coefficient $\rho$,
and~$\operatorname{trace}(\cdot)$
returns the sum of main diagonal elements of its matrix argument.
The $(i,j)$-th entry of $\mathbf{R}_{\mathbf{xx}}$
is given by
$\rho^{|i-j|}$,
$i,j=0,1,\ldots,7$~\cite{Britanak2007, Ahmed1975}.
The correlation coefficient
is assumed as equal to 0.95,
which is representative
for natural images~\cite{Britanak2007}.
\color{black}

\subsubsection{Unified Transform Coding Gain}

Unified transform coding gain provides a measure
to quantify the compression capabilities
of a given matrix~\cite{Tablada2015}.
It is a generalization of
usual transform coding gain as in Reference~\cite{Britanak2007}.
Let~$\mathbf{g}_k$ and~$\mathbf{h}_k$
be the $k$th row of
$\hat{\mathbf{C}}_{\pmb{\alpha}}^{\top}$
and~$\mathbf{C}_{\pmb{\alpha}}$, respectively.
Then,
the unified transform coding gain is given by Reference~\cite{Katto1991}:
\begin{align}
C_g(\hat{\mathbf{C}}_{\pmb{\alpha}})
&
=
10
\cdot
\operatorname{log}_{10}
\left[
\prod_{k = 1}^{8}
\frac{1}{(A_k \cdot B_k)^\frac{1}{8}}
\right]
\quad
\text{(dB)}
,
\end{align}
where
$A_k
=
\operatorname{sum}
[
(\mathbf{h}_k\cdot \mathbf{h}_k^{\top})\circ\mathbf{R}_{\mathbf{xx}}
]$,
$\operatorname{sum}(\cdot)$
returns the sum of elements of its matrix argument,
 operator~$\circ$ denotes the element-wise matrix product,
$B_k = \|\mathbf{g}_k\|_2^2$,
and~$\|\cdot\|_2$ is the usual vector norm.

\subsubsection{Transform Efficiency}

Another measure for assessing
 coding performance is transform efficiency~\cite{Britanak2007}.
Let matrix
$
\mathbf{R}_{\mathbf{XX}}
=
\hat{\mathbf{C}}_{\pmb{\alpha}}
\cdot
\mathbf{R}_{\mathbf{xx}}
\cdot
\hat{\mathbf{C}}_{\pmb{\alpha}}^\top
$
be the covariance matrix of transformed signal~$\mathbf{X}$.
The transform efficiency of
$\hat{\mathbf{C}}_{\pmb{\alpha}}$ is given by~\cite{Britanak2007}:
\begin{align}
\eta
\left(
\hat{\mathbf{C}}_{\pmb{\alpha}}
\right)
&
=
\frac
{\operatorname{trace}\left(|\mathbf{R}_{\mathbf{XX}}|\right)}
{\operatorname{sum}\left(|\mathbf{R}_{\mathbf{XX}}|\right)}
.
\end{align}

\subsection{Computational Cost}
\label{subsec:computational_cost}

\begin{figure}
\centering
\includegraphics[scale=1]{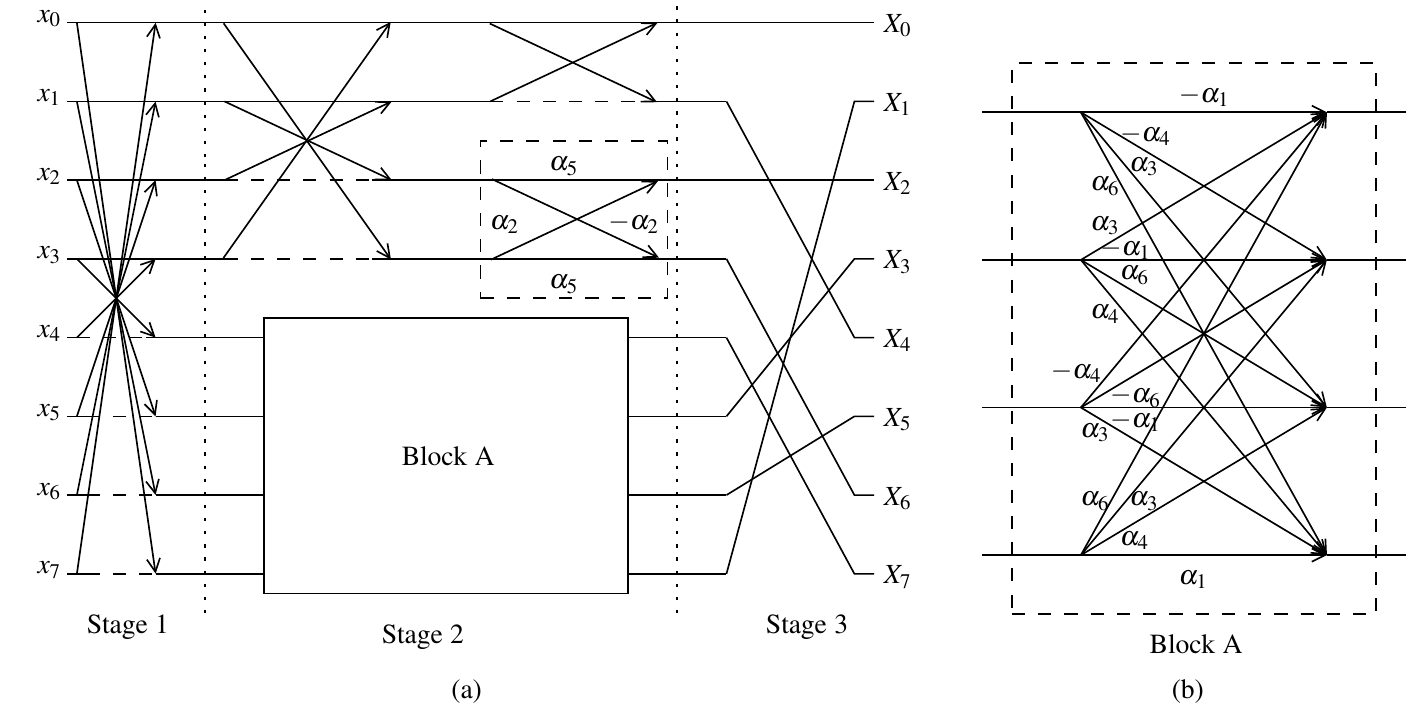}
\caption{Signal flow graph (SFG) of the Loeffler-based transformations.
Dashed lines represents multiplication by~$-1$.
Stages~1, 2, and~3 are represented by matrices~$\mathbf{A}, \mathbf{M'},$ and~$\mathbf{P}$, respectively, as in Matrix~\eqref{eq:dct-factorization}.} %

\label{ApprSFG}

\end{figure}

Based on Loeffler DCT factorization,
a fast algorithm for $\mathbf{T}_{\pmb{\alpha}}$
is obtained and its signal flow graph (SFG) is shown in Figure~\ref{ApprSFG}.
Stage 1,~2, and 3
correspond
to matrices
$\mathbf{A}$,
$\mathbf{M}_{\pmb{\alpha}}$,
and
$\mathbf{P}$,
respectively.
The computational cost
of such an algorithm
is closely linked to selected parameter values
$\alpha_k$,
\linebreak $k=1,2,\ldots,6$.
Because we aim at proposing
multiplierless approximations,
we restricted parameter values $\alpha_k$
to set
$\mathcal{P} = \{0, \pm 1/2, \pm 1, \pm 2\}$,
i.e.,
$\pmb{\alpha} \in \mathcal{P}^6$.
The elements in~$\mathcal{P}$
correspond to trivial multiplications
that affect null multiplicative complexity.
In fact,
the elements in $\mathcal{P}$
represent only additions and minimal bit-shifting operations.

The number of additions and bit-shifts can be
evaluated by inspecting
the discussed algorithm (Figure~\ref{ApprSFG}).
Thus,
we obtain the following expressions
for the addition and bit-shifting counts,~respectively:
\begin{equation}
\begin{split}
A(\pmb{\alpha})
=
&
8
+
2
\cdot
\operatorname{max}
\Big\{
1,
\mathbf{1}_{\mathcal{P}-\{0\}}(\alpha_2)
+
\mathbf{1}_{\mathcal{P}-\{0\}}(\alpha_5)
\Big\}
\\
&
+
4
\cdot
\operatorname{max}
\left\{
1,
\sum_{k\in\{1,3,4,6\}}
\mathbf{1}_{\mathcal{P}-\{0\}}(\alpha_k)
\right\}
,
\end{split}
\end{equation}
\begin{equation}
\begin{split}
S(\pmb{\alpha})
=
&
2
\cdot
[
\mathbf{1}_{\{\pm\frac{1}{2}, \pm2\}}(\alpha_2)
+
\mathbf{1}_{\{\pm\frac{1}{2}, \pm2\}}(\alpha_5)
]
\\
&
+
4
\cdot
\sum_{k\in\{1,3,4,6\}}
\mathbf{1}_{\{\pm\frac{1}{2}, \pm2\}}(\alpha_k)
,
\end{split}
\end{equation}
where
$\mathbf{1}_A(x) = 1$, if $x\in A$,
and 0 otherwise.

\section{Multicriteria Optimization and New Transforms}
\label{multicriteriaoptimization}

In this section,
we introduce an optimization problem
that aims at identifying optimal
transformations derived from the proposed mapping (Matrix~\eqref{matrixmapping}). %
Considering the various
performances and complexity metrics
discussed in the previous section,
we set up the following
multicriteria optimization problem~\cite{Ehrgott2005, Barichard2009}:
\begin{align}
\label{optm}
\underset
{{\pmb{\alpha}}\in \mathcal{P}^6}
{\operatorname{min}}
\left(
\epsilon(\hat{\mathbf{C}}_{\pmb{\alpha}}),
\operatorname{MSE}(\hat{\mathbf{C}}_{\pmb{\alpha}}),
-C_g(\hat{\mathbf{C}}_{\pmb{\alpha}}),
-\eta(\hat{\mathbf{C}}_{\pmb{\alpha}}),
A(\pmb{\alpha}),
S(\pmb{\alpha})
\right)
,
\end{align}
subject to:
\begin{enumerate}[align=parleft,leftmargin=*,labelsep=3mm]

\item[i]
the
existence of inverse transformation,
according to the condition established in Matrix~\eqref{detU}; %

\item[ii]
the entries of the inverse matrix
must be in $\mathcal{P}$;
to ensure both forwarded and inverse low-complexity transformations;

\item[iii]
the property of
orthogonality or near-orthogonality
according to the criterion in Equation~\eqref{equation-nearly-orthogonality-criterion}. %

\end{enumerate}
Quantities~$C_g(\hat{\mathbf{C}}_{\pmb{\alpha}})$
and
$\eta(\hat{\mathbf{C}}_{\pmb{\alpha}})$
are in negative form
to comply to the minimization requirement.

Being a multicriteria optimization problem,
Problem \eqref{optm} is based on objective function
set
$\mathcal{F} = \{\epsilon(\cdot), \operatorname{MSE}(\cdot), -C_g(\cdot), -\eta(\cdot), A(\cdot), S(\cdot)\}$.
The problem in analysis is discrete and finite since there is a countable number of values to the objective function.
However,
the nonlinear, discrete nature of the problem
renders it
unsuitable for analytical methods.
Therefore,
we employed exhaustive search methods
to solve it.
The discussed multicriteria problem
requires the identification
of the Pareto efficient solutions set~\cite{Ehrgott2005},
which is given by:
\begin{equation}
\begin{split}
\{\pmb{\alpha^*} \in \mathcal{P}^6:&
\text{ there is no $\alpha \in \mathcal{P}^6$, such that} \\
& f(\pmb{\alpha}) \leq f(\pmb{\alpha^*}) \text{ for all $f \in \mathcal{F}$ and}\\
& f_0(\pmb{\alpha}) < f_0(\pmb{\alpha^*}) \text{ for some $f_0 \in \mathcal{F}$}
\}
.
\end{split}
\end{equation}

\subsection{Efficient Solutions}

The exhaustive search~\cite{Ehrgott2005} returned
six efficient
parameter vectors,
which are listed in
Table~\ref{effsolpar}.
For~ease of notation,
we denote
the low-complexity matrices and their associated approximations
linked to efficient solutions
according to:
$
\mathbf{T}_i
\triangleq
\mathbf{T}_{\pmb{\alpha}_i^\ast}
$
and
$
\hat{\mathbf{C}}_i
\triangleq
\hat{\mathbf{C}}_{\pmb{\alpha}_i^\ast}
$,
respectively.
Table~\ref{effsolm}
summarizes
the performance metrics,
arithmetic complexity,
and
orthonormality property
of obtained
matrices
$\hat{\mathbf{C}}_i$,
$i = 1, 2, \ldots, 6$.
We included the DCT for reference as well.
Note that all DCT approximations
except those by~$\hat{\mathbf{C}}_3$
are orthonormal.

\begin{table}
\centering
\caption{Efficient solutions.}
\label{effsolpar}
\begin{tabular}{c@{\qquad}c@{\qquad}}\toprule
\boldmath{$i$} & \boldmath{$\alpha_i^*$}\\ \midrule
$1$ & $[ 1  \quad 1  \quad 0   \quad 0  \quad 0  \quad 0   ]^{\top}$\\
$2$ & $[ 1  \quad 1  \quad 0  \quad 0  \quad \frac{1}{2}  \quad 0   ]^{\top}$\\
$3$ & $[ 1  \quad 1  \quad 1   \quad 0  \quad 0  \quad 0   ]^{\top}$\\
$4$ & $[ 1  \quad 1  \quad 1   \quad 1  \quad \frac{1}{2}  \quad 0   ]^{\top}$\\
$5$ & $[ 1  \quad 2  \quad 0   \quad 0  \quad 1  \quad 0   ]^{\top}$\\
$6$ & $[ 1  \quad 2  \quad 1   \quad 1  \quad 1  \quad 0   ]^{\top}$\\ \bottomrule

\end{tabular}
\end{table}
\unskip

\begin{table}
\centering

\caption{Efficient Loeffler-based discrete cosine transform (DCT) approximations and the DCT.}
\label{effsolm}
\scalebox{0.9}[0.9]{
\begin{tabular}{c@{\:\:\:}c@{\:\:\:}c@{\:\:\:}c@{\:\:\:}c@{\:\:\:}c@{\:\:\:}c@{\:\:\:}c@{\:\:\:}l@{\:\:\:}}\toprule
\textbf{Transform} &
\boldmath{$\epsilon$}&
\boldmath{$\operatorname{MSE}$} &
\boldmath{$C_g$} &
\boldmath{$\eta$} &
\boldmath{$A(\pmb{\alpha})$} &
\boldmath{$S(\pmb{\alpha})$} &

\textbf{Orthonormal?} &
\textbf{Description}\\ \midrule
$\hat{\mathbf{C}}_1$ & $8.66$ & $0.059$ & $7.33$ & $80.90$& $14$ & $0$& Yes & Proposed in Reference~\cite{Cintra2012}\\
$\hat{\mathbf{C}}_2$ & $7.73$ & $0.056$ & $7.54$ & $81.99$ & $16$ & $2$& Yes & Proposed as~$\mathbf{T}_{10}$ in Reference~\cite{Tablada2015} \\%(denoted~$\mathbf{T}_{10}$ in~\cite{Tablada2015})\\
$\hat{\mathbf{C}}_3$ & $1.44$ &$0.007$ & $8.30$ & $89.77$ & $18$ & $0$ & No &  Proposed as~$\tilde{\mathbf{T}}_1$ in Reference~\cite{Cintra2014} \\%(denoted~$\tilde{\mathbf{T}}_1$ in~\cite{Cintra2014})\\
$\hat{\mathbf{C}}_4$ &   $0.87$ & $0.006$ & $8.39$ & $88.70$ & $24$&$2$& Yes & Proposed as~$\hat{\mathbf{D}}_1$ in Reference~\cite{Lengwehasatit2004} \\%(denoted~$\hat{\mathbf{D}}_1$ in~\cite{Lengwehasatit2004})\\
$\hat{\mathbf{C}}_5$ & $7.73$ & $0.056$ & $7.54$ & $81.99$ & $16$ & $2$& Yes & Equivalent to~$\hat{\mathbf{C}}_2$ and proposed as~$\mathbf{T}_{4}$ in Reference~\cite{Tablada2015} \\%(denoted~$\mathbf{T}_{4}$ in~\cite{Tablada2015})\\
$\hat{\mathbf{C}}_6$ &   $0.87$ & $0.006$ & $8.39$ & $88.70$ & $24$&$2$& Yes & Equivalent to~$\hat{\mathbf{C}}_4$ and proposed as~$\mathbf{T}_{9}$ in Reference~\cite{Tablada2015} \\

$\text{DCT}$ &  0 & 0 & $8.85$ & $93.99$ & -- & -- & Yes &  Exact DCT as described in Reference~\cite{Britanak2007}\color{black}\\\bottomrule

\end{tabular}}
\end{table}

\subsection{Comparison}

Several DCT approximations
are encompassed by the proposed matrix formalism.
Such transformations include:
the SDCT~\cite{Haweel2001},
the approximation based on round-off function proposed in Reference~\cite{Cintra2011a},
and
all the DCT approximations introduced in Reference~\cite{Cintra2014}.
For instance,
the SDCT~\cite{Haweel2001}
is another particular transformation fully described by
the proposed matrix mapping.
In fact,
the SDCT can be obtained
by taking
$f(\pmb{\alpha}_1)$,
where
$\pmb{\alpha}_1 = \begin{bmatrix}1&1&1&1&1&1\end{bmatrix}^\top$.
Nevertheless,
none of these approximations
is part of the Pareto efficient solution set
induced
by the discussed multicriteria optimization problem~\cite{Ehrgott2005}.
Therefore,
we compare the obtained efficient solutions
with a variety of state-of-the-art
eight-point DCT approximations
that cannot be described by the proposed Loeffler-based formalism.
We separated
the Walsh--Hadamard transform (WHT)
and
the Bouguezel--Ahmad--Swamy (BAS) series of approximations
labeled
$\text{BAS}_1$~\cite{Bouguezel2008}, $\text{BAS}_2$~\cite{Bouguezel2008a},
$\text{BAS}_3$~\cite{Bouguezel2009}, $\text{BAS}_4$~\cite{Bouguezel2010},
$\text{BAS}_5$~\cite{Bouguezel2011} \mbox{(for $a = 1$)},
$\text{BAS}_6$~\cite{Bouguezel2011} \mbox{(for $a = 0$)},
$\text{BAS}_7$~\cite{Bouguezel2011} \mbox{(for $a = 1/2$)},
and
$\text{BAS}_8$~\cite{Bouguezel2013}.
Table~\ref{BASm} shows the performance measures for these transforms.
For completeness, we also show the unified coding gain and the transform efficiency measures for the exact DCT~\cite{Britanak2007}.

Some approximations, such as the SDCT, were not explicitly included in our comparisons.
Although they are in the set of matrices generated by Loeffler parametrization, they are not in the efficient solution set.
Thus, we removed them from further analyses for not being an optimal solution.

\begin{table}
\centering

\scriptsize
\caption{Performance of the Bouguezel--Ahmad--Swamy (BAS) approximations, the Walsh--Hadamard transform (WHT), and the DCT.}
\label{BASm}
\begin{tabular}{c@{\:\:\:}c@{\:\:\:}c@{\:\:\:}c@{\:\:\:}c@{\:\:\:}c@{\:\:\:}c@{\:\:\:}c@{\:\:\:}}
\toprule
\textbf{Transform} &
\textbf{Orthogonalizable} &

\boldmath{$\epsilon$}&
\boldmath{$\operatorname{MSE}$} &
\boldmath{$C_g$} &
\boldmath{$\eta$} &

\boldmath{$A(\pmb{\alpha})$} &
\boldmath{$S(\pmb{\alpha})$}
\\
\midrule
$\text{BAS}_1$~\cite{Bouguezel2008} & No & $4.19$ & $0.019$ & $6.27$ & $83.17$& $21$ & $0$\\
$\text{BAS}_2$~\cite{Bouguezel2008a} & Yes & $5.93$ & $0.024$ & $8.12$ & $86.86$ & $18$ & $2$\\
$\text{BAS}_3$~\cite{Bouguezel2009} & Yes & $6.85$ &$0.028$ & $7.91$ & $85.38$ & $18$ & $0$ \\
$\text{BAS}_4$~\cite{Bouguezel2010} & Yes & $4.09$ & $0.021$ & $8.33$ & $88.22$ & $24$ & $4$\\
$\text{BAS}_5$~\cite{Bouguezel2011} & Yes & $26.86$ & $0.071$ & $7.91$ & $85.38$ & $18$ & $0$\\
$\text{BAS}_6$~\cite{Bouguezel2011} & Yes & $26.86$ & $0.071$ & $7.91$ & $85.64$ & $16$ & $0$\\
$\text{BAS}_7$~\cite{Bouguezel2011} & Yes & $26.40$ & $0.068$ & $8.12$ & $86.86$ & $18$ & $2$\\
$\text{BAS}_8$~\cite{Bouguezel2013} & Yes & $35.06$ & $0.102$ & $7.95$ & $85.31$ & $24$ & $0$\\
$\text{WHT}$~\cite{Britanak2007} & Yes & $5.05$ & $0.025$ & $7.95$ & $85.31$ & $24$ & $0$\\
$\text{DCT}$~\cite{Britanak2007} & Yes & 0 & 0 & $8.85$ & $93.99$ & -- & --\\\bottomrule
\end{tabular}
\end{table}

\begin{figure}
\centering
\includegraphics[scale=0.88]{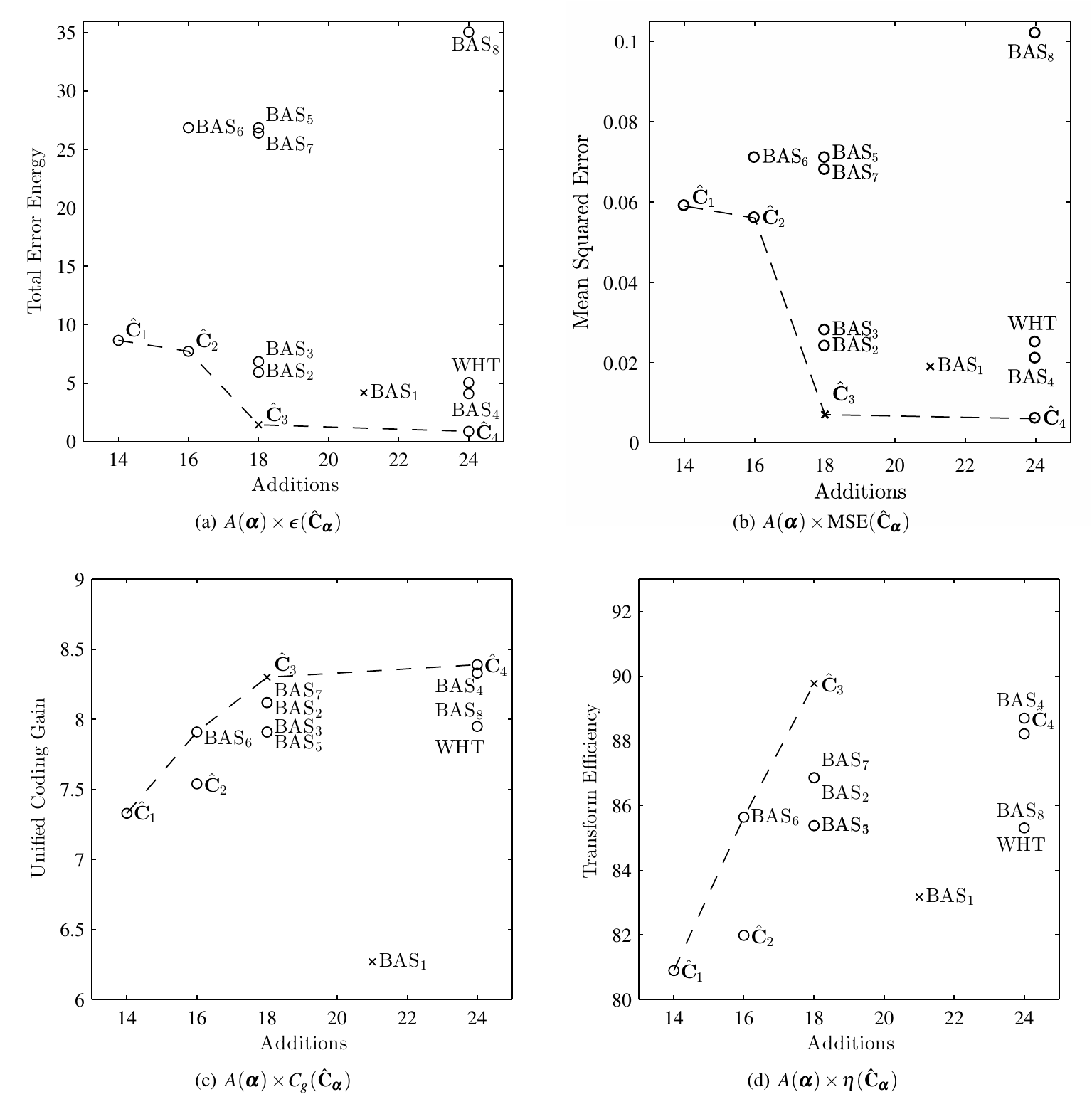}
\caption{Performance plots and Pareto frontiers
for the discussed transformations.
Orthogonal transforms are marked with circles ($\circ$)
and nonorthogonal approximations with
cross sign ($\times$).
Dashed curves represent
the~Pareto frontier.
}
\label{paretolines}
\end{figure}

In order to compare
all the above-mentioned transformations,
we aimed at
identifying
the
Pareto frontiers~\cite{Ehrgott2005}
in
two-dimensional plots
considering
the performance figures
of
the obtained efficient solution
as well as
the WHT and BAS approximations.
Thus,
we devised scatter plots
considering the
arithmetic complexity
and
performance measures.
The resulting plots are shown in Figure~\ref{paretolines}.
Orthogonal transform approximations are marked with circles,
and nonorthogonal approximations with cross signs.
The dashed curves represent
the~Pareto frontier~\cite{Ehrgott2005}
for each selected pair of the measures.
Transformations located on the Pareto frontier
are considered optimal,
where the points are dominated
by the frontier
correspond to nonoptimal transformations.
The bivariate plots
in Figure~\ref{paretolines}a,b
reveal
that the obtained Loeffler-based
DCT approximations
are often situated
at the optimality site
prescribed by the Pareto frontier.
The Loeffler approximations
perform particularly well
in terms of
 total error energy
and
the MSE,
which
capture the matrix proximity
to the exact DCT matrix
in a Euclidean sense.
Such approximations are particularly
suitable for problems that require
computational proximity to the the exact transformation
as in the case of detection and estimation problems~\cite{Kay1993,Kay1998}.
Regarding coding performance,
Figure~\ref{paretolines}c,d
shows that
transformations
$\hat{\mathbf{C}}_1$,
$\hat{\mathbf{C}}_3$,
$\hat{\mathbf{C}}_6$,
and
$\text{BAS}_6$
are situated on the  Pareto frontier,
being optimal in this sense.
These approximations are adequate for data compression
and decorrelation~\cite{Britanak2007}.

\section{Image and Video Experiments}
\label{sec:experiments}
\vspace{-6pt}

\subsection{Image Compression}

\begin{figure}
\centering
\includegraphics[scale=0.9]{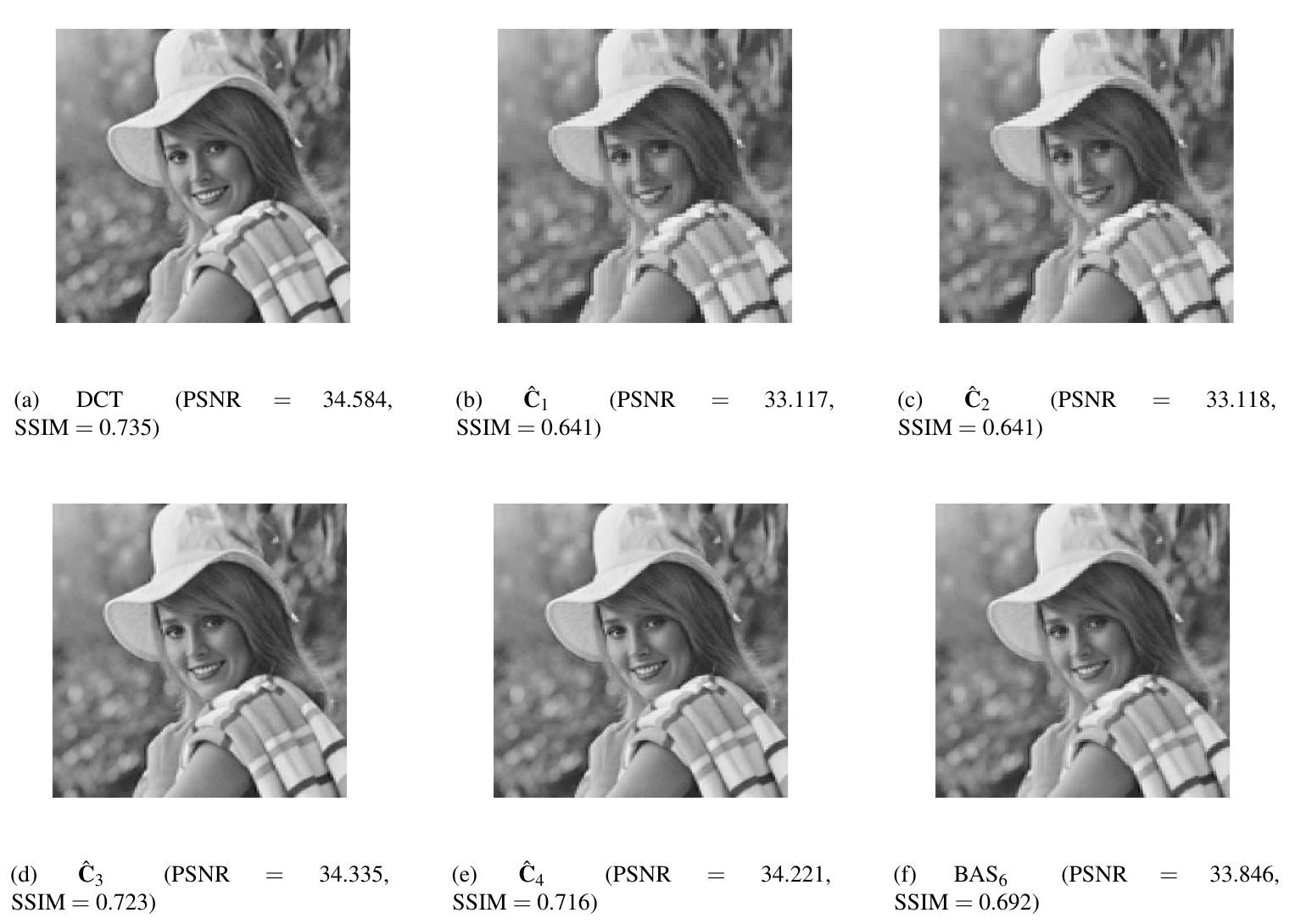}
\caption{Elaine image compressed for DCT and selected approximations:
$\hat{\mathbf{C}}_1$,
$\hat{\mathbf{C}}_2$,
$\hat{\mathbf{C}}_3$
$\hat{\mathbf{C}}_4$,
and
$\text{BAS}_6$.
Only five transform-domain coefficients were retained.}

\label{elaine}

\end{figure}

We implemented the JPEG-like compression experiment
described in References~\cite{Bouguezel2008,Bouguezel2009,Bouguezel2011,Bouguezel2011,Haweel2001}
and submitted
the standard Elaine image to processing
at a high compression rate.
For quantitative assessment,
we adopted the structural similarity (SSIM) index~\cite{Wang2004}
and the
peak signal-to-noise rate (PSNR)~\cite{Bhaskaran1995, Gonzalez2001} measure.
Figure~\ref{elaine} shows
the
reconstructed Elaine image
according to the JPEG-like compression
considering the following transformations:
DCT,
$\hat{\mathbf{C}}_1$,
$\hat{\mathbf{C}}_2$,
$\hat{\mathbf{C}}_3$,
$\hat{\mathbf{C}}_4$,
and
$\text{BAS}_6$.
We employed fixed-rate compression
and retained only five coefficients,
which led to 92.1875\% compression rate.
Despite very low computational complexity,
the approximations could furnish
images with quality comparable
to the results obtained from the exact DCT.
In particular,
approximations
$\hat{\mathbf{C}}_1$ and
$\hat{\mathbf{C}}_2$
offered good trade-off,
since they required only 14--16 additions
and
were capable of providing
competitive image quality
at smaller hardware and power requirements (Section~\ref{sec:fpga}).

\color{black}

\begin{figure}
\centering
\psfrag{C1}[][][0.8]{$\hat{\mathbf{C}}_1$}
\psfrag{C2}[][][0.8]{$\hat{\mathbf{C}}_2$}
\psfrag{C3}[][][0.8]{$\hat{\mathbf{C}}_3$}
\psfrag{C4}[][][0.8]{$\hat{\mathbf{C}}_4$}
\psfrag{BAS6}[][][0.8]{BAS$_6$}
\psfrag{DCT}[][][0.8]{DCT}
\psfrag{PSNR (dB)}[][][0.8]{PSNR (dB)}
\psfrag{Retained Coefficients}[][][0.8]{Retained Coefficients}
\subfigure[]{\label{elaine_r.psnr}\includegraphics[scale=0.9]{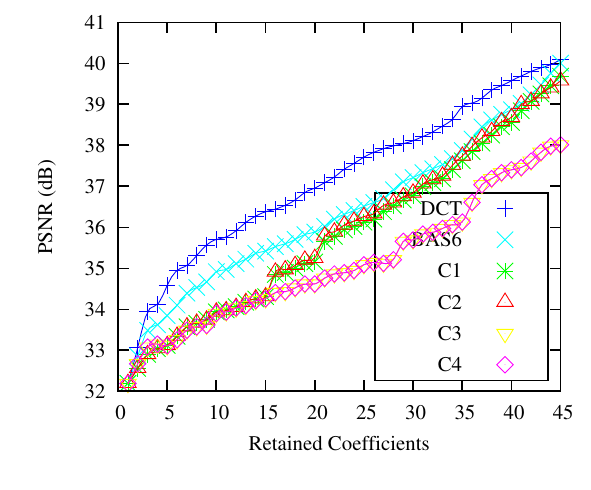}
}
~
\subfigure[]{\label{elaine_r.ssim}\includegraphics[scale=0.9]{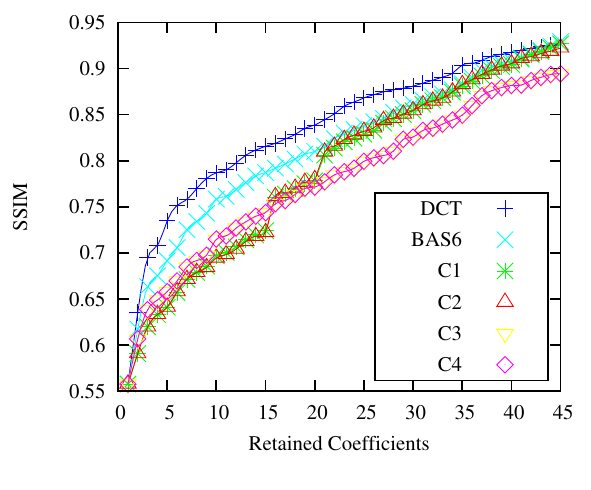}
}
\caption{Peak signal-to-noise rate (PSNR) and structural similarity (SSIM) for the Elaine image compressed for DCT, and selected approximations
$\hat{\mathbf{C}}_1$,
$\hat{\mathbf{C}}_2$,
$\hat{\mathbf{C}}_3$
$\hat{\mathbf{C}}_4$,
and
$\text{BAS}_6$.}

\label{elaine_r}

\end{figure}

Figure~\ref{elaine_r}
shows the PSNR and SSIM
for different number of
retained coefficients.
Considering PSNR measurements,
the difference
between the measurements
associated to the $\text{BAS}_6$ and
to the efficient approximation
were
less than $\approx$1~dB.
Similar
behavior was reported
when SSIM measurements were considered.

\color{black}

\subsection{Video Compression}

\vspace{-6pt}

\subsubsection{Experiments with the \mbox{H.264}/AVC Standard}

To assess the Loeffler-based approximations
in the context of
video coding,
we embedded them
in the
x264~\cite{x264team} software library
for encoding video streams into
the \mbox{H.264}/AVC standard~\cite{Richardson2010}.
The original eight-point transform employed in \mbox{H.264}/AVC
is a DCT-based
integer approximation given by Reference~\cite{Gordon2004}:
\begin{align}
\mathbf{\hat{C}}_\text{AVC}
=
\frac{1}{8}
\cdot
\left[
\begin{rsmallmatrix} 8 & 8 & 8 & 8 & 8 & 8 & 8 & 8 \\
       12 & 10 & 6 & 3 & -3 & -6 & -10 & -12 \\
        8 & 4 & -4 & -8 & -8 & -4 & 4 & 8 \\
       10 & -3 & -12 & -6 & 6 & 12 & 3 & -10 \\
        8 & -8 & -8 & 8 & 8 & -8 & -8 & 8 \\
        6 & -12 & 3 & 10 & -10 & -3 & 12 & -6 \\
        4 & -8 & 8 & -4 & -4 & 8 & -8 & 4 \\
        3 & -6 & 10 & -12 & 12 & -10 & 6 & -3 \\
\end{rsmallmatrix}
\right].
\label{eq:intDCT}
\end{align}
The fast algorithm for the above transformation
requires 32~additions and 14~bit-shifting operations~\cite{Gordon2004}.

We encoded 11 common intermediate format~(CIF)
videos with 300 frames
from a public video database~\cite{videos}
using the standard and modified codec.
In our simulation,
we employed default settings
and the resulting video quality
was assessed by means
of
the average PSNR of
chrominance and luminance representation
considering all reconstructed frames.
Psychovisual optimization was also disabled
in order to obtain valid PSNR values
and
the quantization step
was~unaltered.
\color{black}

We computed the Bj{\o}ntegaard delta rate (BD-Rate) and the Bj{\o}ntegaard delta PSNR (BD-PSNR)~\cite{Bjontegaard2001} for modified codec compared to the usual H.264 native DCT approximation.
 BD-Rate and BD-PSNR
were automatically
furnished
by the x264~\cite{x264team} software library.
A comprehensive review of Bj{\o}ntegaard metrics
and
their specific mathematical formulations
can be found in Reference~\cite{Hanhart2014}.
\color{black}
For~that, we followed the procedure specified in References~\cite{Hanhart2014, Bjontegaard2001, Tan2016}.
We adopted the same testing point as determined in Reference~\cite{Bossen2014} with
fixed quantization parameter (QP)~in~$\{22, 27, 32, 37\}$.
Following References~\cite{Hanhart2014,Tan2016}, we employed cubic spline interpolation between the testing points for better visualization.

\begin{figure}
\centering
\includegraphics[scale=1]{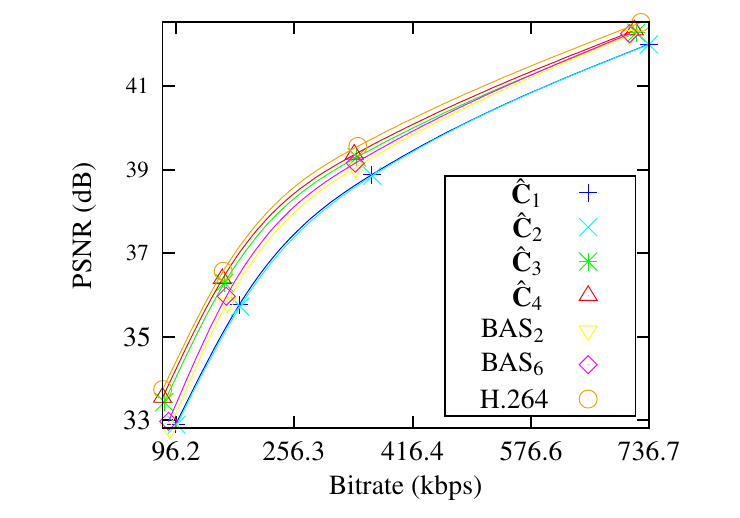}

\caption{
Average \color{black} rate-distortion curves of the modified \mbox{H.264}/AVC software for the selected 11~CIF videos from Reference~\protect\cite{videos}
for the PSNR (dB) in terms of bitrate.
}

\label{figure:bitrate}
\end{figure}

Figure~\ref{figure:bitrate} shows the resulting
average
\color{black}
BD-Rate distortion curves for the selected videos,
where H.264 represents the integer DCT used in the x264 software library shown in Matrix~\eqref{eq:intDCT}. %
We note the superior performance of~$\hat{\mathbf{C}}_3$ and~$\hat{\mathbf{C}}_4$ approximations compared to BAS$_2$ and BAS$_6$ transforms.
This performance is more evident for small bitrates.
As the bitrate increases, the quality performance of~$\hat{\mathbf{C}}_3$, $\hat{\mathbf{C}}_4$, BAS$_2$, and BAS$_6$ approximates native DCT implementation.
Table~\ref{tab.bd-measures-h.264} shows the BD-Rate and BD-PSNR measures
for the 11~CIF video sequence selected from Reference~\cite{videos}.

\begin{figure}
\centering
\includegraphics[scale=0.9]{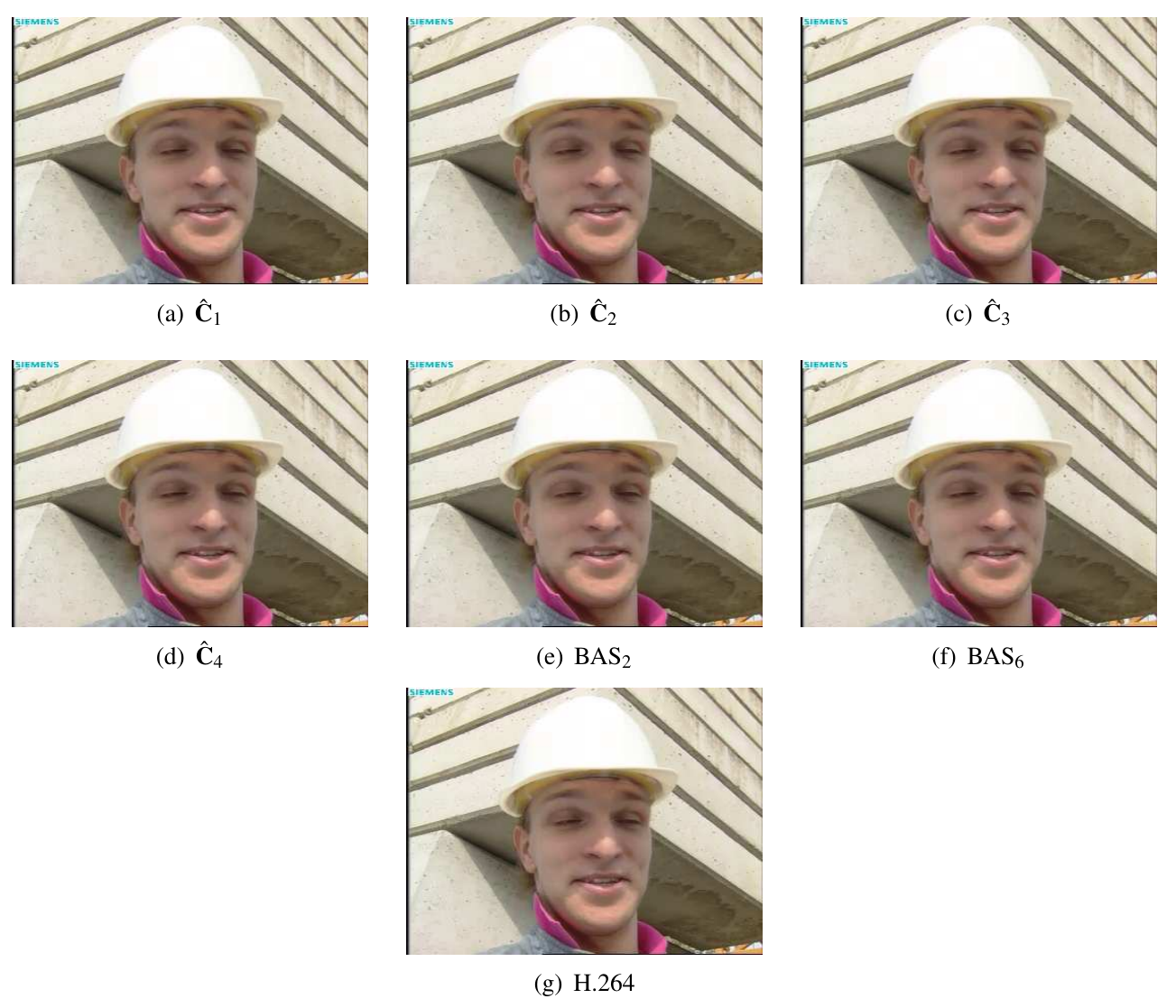}
\caption{First frame of the \mbox{H.264}/AVC compressed `Container' sequence with $\text{QP}=27$.} %
\label{figure:video}
\end{figure}
\unskip
\begin{table}
\centering
\caption{Average Bj{\o}ntegaard delta (BD)-Rate and BD-PSNR for the 11 common intermediate format (CIF) videos selected from Reference~\cite{videos}.}
\label{tab.bd-measures-h.264}
\begin{tabular}{c@{\:\:\:}c@{\:\:\:}c@{\:\:\:}}
\toprule
\textbf{Transform} & \textbf{BD-Rate (\%)} & \textbf{BD-PSNR (dB)}\\\midrule
$\hat{\mathbf{C}}_1$& $29.98$ & $-1.08$ \\
$\hat{\mathbf{C}}_2$& $31.40$ & $-1.13$ \\
$\hat{\mathbf{C}}_3$& $7.48$ & $-0.29$ \\
$\hat{\mathbf{C}}_4$& $3.94$ & $-0.15$  \\
BAS$_2$& $17.43$ & $-0.68$ \\
BAS$_6$& $13.90$ & $-0.54$\\\bottomrule
\end{tabular}
\end{table}
Note that Loeffler approximations~$\hat{\mathbf{C}}_1$ and~$\hat{\mathbf{C}}_2$, again, show better performance in terms of BD-PSNR measures than BAS transforms~BAS$_2$ and~BAS$_6$.
We attribute the difference in quality between~$\hat{\mathbf{C}}_3$ and~$\hat{\mathbf{C}}_4$ in relation to~$\hat{\mathbf{C}}_1$ and~$\hat{\mathbf{C}}_2$ to the associated complexity discrepancy as depicted in Table~\ref{effsolm}.
The reduction in the number of additions and bit-shifting operations come at the expense of performance.

\subsubsection{Experiments with the \mbox{H.265}/HEVC Standard}

We also embedded Loeffler-based approximations
into the \mbox{HM-16.3}
\mbox{H.265}/HEVC reference software~\cite{hmteam}.
The \mbox{H.265}/HEVC employs the following integer transform:
\begin{align}
\mathbf{\hat{C}}_\text{HEVC}
=
\frac{1}{64}
\cdot
\left[
\begin{rsmallmatrix}
64 &64 &64 &64 &64 &64 &64 &64 \\
89 &75 &50 &18 &-18 &-50 &-75 &-89 \\
83 &36 &-36 &-83 &-83 &-36 &36 &83 \\
75 &-18 &-89 &-50 &50 &89 &18 &-75 \\
64 &-64 &-64 &64 &64 &-64 &-64 &64 \\
50 &-89 &18 &75 &-75 &-18 &89 &-50 \\
36 &-83 &83 &-36 &-36 &83 &-83 &36 \\
18 &-50 &75 &-89 &89 &-75 &50 &-18
\end{rsmallmatrix}
\right],
\label{eq:intDCT_hevc}
\end{align}
which demands~72 additions and~62 bit-shifting operations,
when considering the factorization  method based on odd and even decomposition described in Reference~\cite{Britanak2007}.
\color{black}
Besides a particular eight-point integer transformation,
the HEVC standard also employs
four-, 16-, and 32-point transforms~\cite{Pourazad2012}.
To obtain a modified version of the \mbox{H.265}/HEVC codec,
we derived 16- and 32-point approximations based on
the scalable recursive method proposed in Reference~\cite{Jridi2015}.
The eight-point Loeffler-based efficient solutions
were employed as fundamental building blocks
for scaling up.
\color{black}
Since the four-point DCT matrix  only possesses
 integer coefficients,
it was employed unaltered.
\color{black}
The additive complexity of the scaled 16- and 32-point approximations
is
$2 \cdot A(\pmb{\alpha}) + 16$
and
$4 \cdot A(\pmb{\alpha}) + 64$,
respectively~\cite{Jridi2015}.
For the bit-shifting count,
we have
$2\cdot S(\pmb{\alpha})$
and
$4\cdot S(\pmb{\alpha})$,
respectively.

In order to perform image-quality assessment,
we encoded the first 100~frames of
one standard sequence of each~A~to~F video classes defined in
the Common Test Conditions
and
Software Reference Configurations (CTCSRC)
document~\cite{CTConditions2013}.
The selected videos and their attributes are made explicit in the first two columns of Table~\ref{tab:videos}.
Classes~A and~B present high-resolution content for broadcasting and video on demand services; Classes~C and~D are lower-resolution contents for the same applications; Class~E reflects typical examples of video-conferencing data; and Class~F provides computer-generated image sequences~\cite{Naccari2014}.
All encoding parameters, including QP values, were set according to the CTCSRC document for the Main profile and All-Intra (AI),
Random Access (RA), Low Delay B (LD-B), and Low Delay P (LD-P) configurations.
In the AI~configuration,
each frame from the input video sequence is independently coded.
RA and LD (B and P) coding modes differ mainly by
order of coding and outputting.
In the former,
the coding order and output order of
the pictures (frames)
may differ, whereas
equal order is required for the latter~(Reference \cite{Wien2015}, p.~94).
RA configuration relates to broadcasting and streaming use,
whereas LD (B and P) represent conventional applications.

It is worth mentioning that Class~A test sequences are not used for
LDB and LDP coding tests,
whereas Class~E videos are not included
in the RA test cases because of the nature of the content
they represent~(Reference \cite{Wien2015}, p.~93).

\begin{figure}
\centering

\includegraphics[scale=0.9]{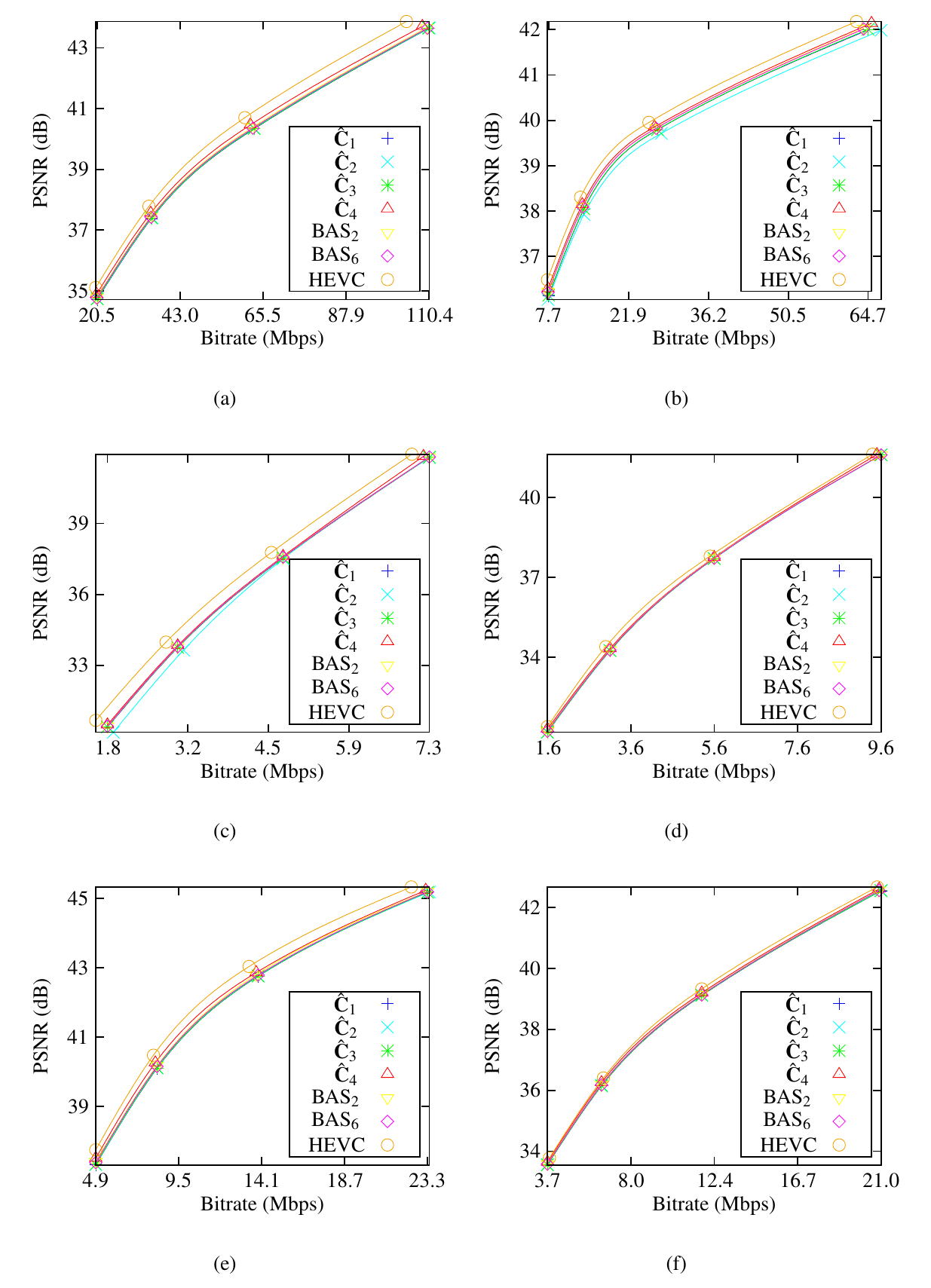}

\caption{%
Rate-distortion curves of the modified HEVC software in AI mode for test sequences:
(\textbf{a})~`PeopleOnStreet',
(\textbf{b})~`BasketballDrive',
(\textbf{c})~`RaceHorses',
(\textbf{d})~`BlowingBubbles',
(\textbf{e})~`KristenAndSara',
and
(\textbf{f})~`BasketbalDrillText’.
}
\label{fig:bdfig6_AI}
\end{figure}

\begin{figure}
\centering
\includegraphics[scale=1]{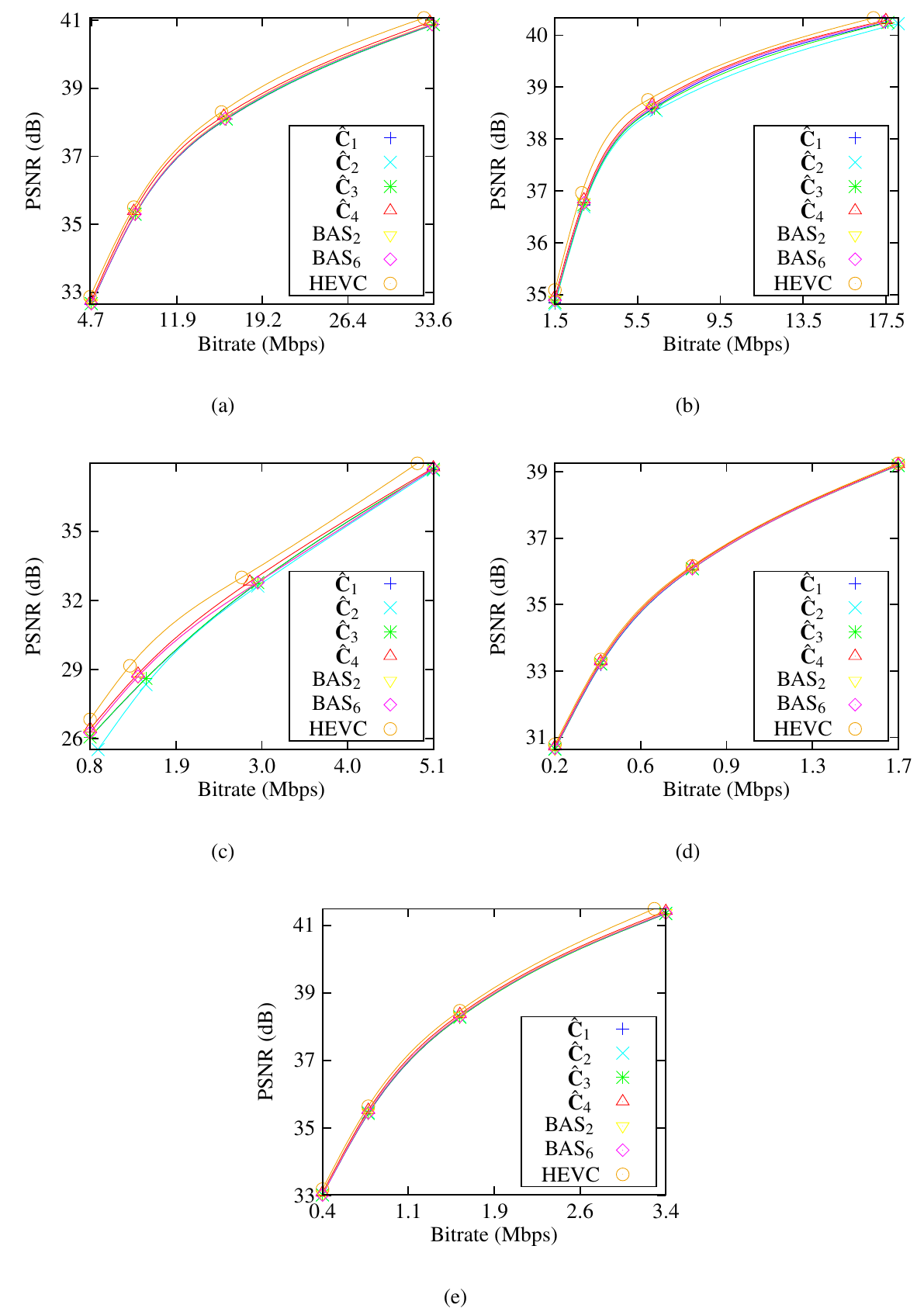}

\caption{%
Rate-distortion curves of the modified HEVC software in RA mode for test sequences:
(\textbf{a})~`PeopleOnStreet',
(\textbf{b})~`BasketballDrive',
(\textbf{c})~`RaceHorses',
(\textbf{d})~`BlowingBubbles',
and
(\textbf{e})~`BasketbalDrillText'.
}
\label{fig:bdfig6_RA}
\end{figure}

\begin{figure}
\centering
\includegraphics[scale=1]{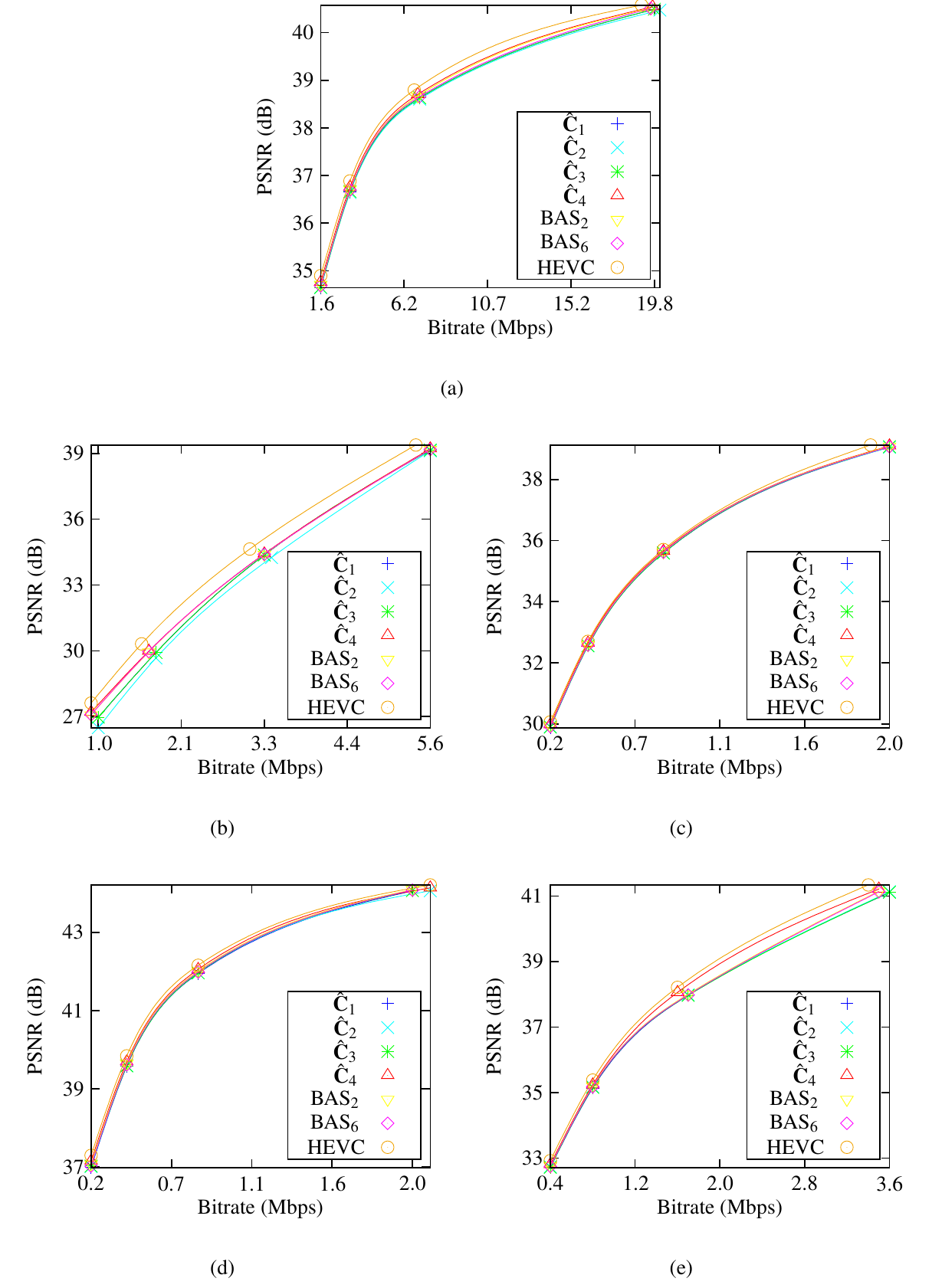}

\caption{%
Rate-distortion curves of the modified HEVC software in LDB mode for test sequences:
(\textbf{a})~`BasketballDrive',
(\textbf{b})~`RaceHorses',
(\textbf{c})~`BlowingBubbles',
(\textbf{d})~`KristenAndSara',
and
(\textbf{e})~`BasketbalDrillText'.
}
\label{fig:bdfig6_LDB}
\end{figure}

\begin{figure}
\centering
\includegraphics[scale=1]{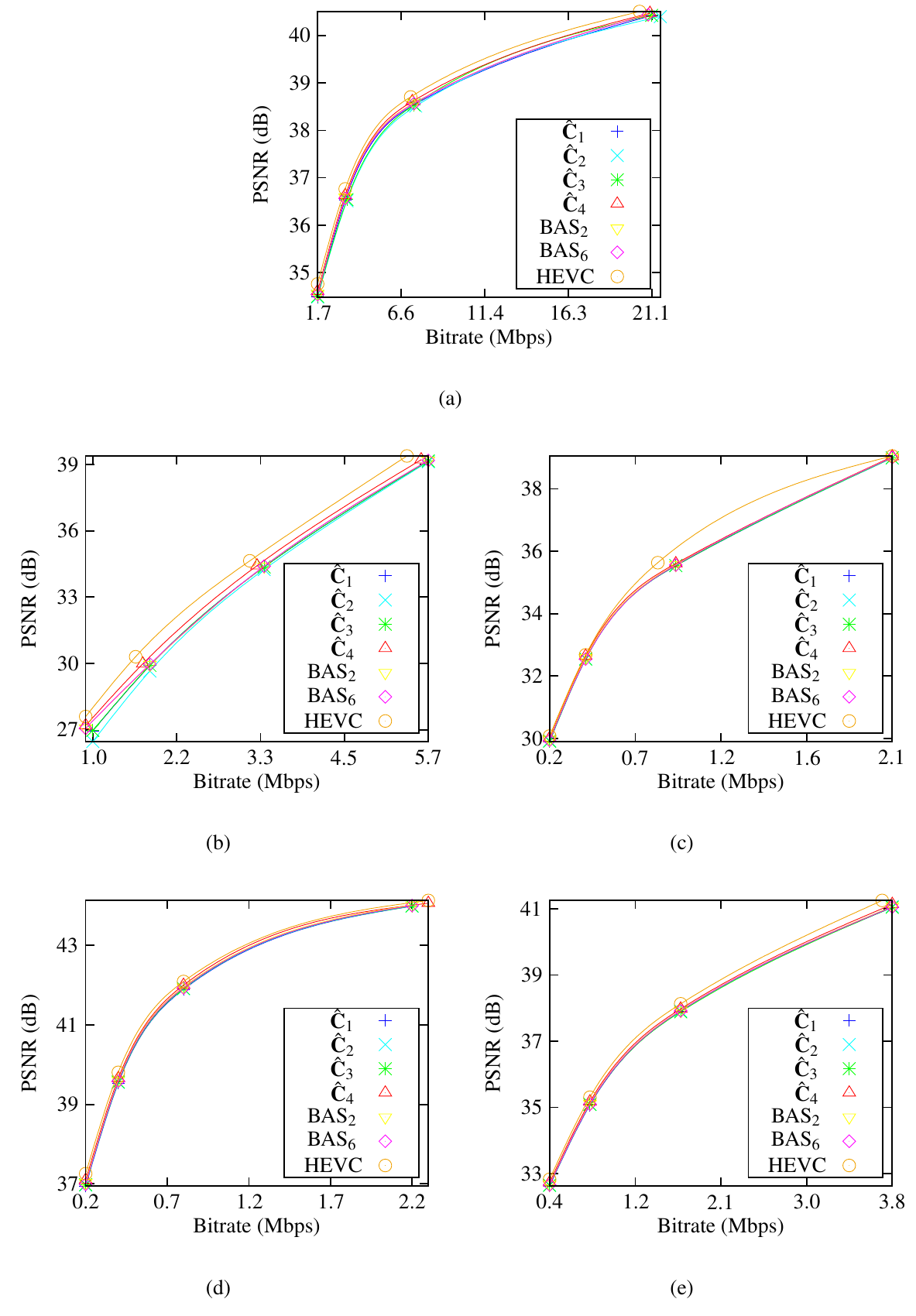}

\caption{%
Rate-distortion curves of the modified HEVC software in LDP mode for test sequences:
(\textbf{a})~`BasketballDrive',
(\textbf{b})~`RaceHorses',
(\textbf{c})~`BlowingBubbles',
(\textbf{d})~`KristenAndSara',
and
(\textbf{e})~`BasketbalDrillText'.
}
\label{fig:bdfig6_LDP}
\end{figure}

As figures of merit, we computed the BD-Rate and BD-PSNR
for the modified versions of the codec.
Figures~\ref{fig:bdfig6_AI}--\ref{fig:bdfig6_LDP}
depict the rate-distortion curves and
Table~\ref{tab:videos}
 presents the Bj{\o}ntegaard metrics for
the considered sets of eight- to 32-point approximations
for modes AI, RA, LDB, and LDP.
 BD-Rate and BD-PSNR are also automatically furnished
by
\mbox{HM-16.3}~\mbox{H.265}/HEVC reference software~\cite{hmteam}.
\color{black}
We employed the cubic spline interpolation based on four testing points
on Figures~\ref{fig:bdfig6_AI}--\ref{fig:bdfig6_LDP},
as specified in References~\cite{Hanhart2014,Tan2016}.
The curves denoted by~HEVC represents the native integer approximations for the eight-, 16-, and 32-point DCT in the
HM-16.3 H.265/HEVC software~\cite{hmteam}.
These four testing points were determined by the specification for HEVC as in Reference~\cite{Bossen2014} with fixed $\text{QP} \in \{ 22, 27, 32, 37\}$.
One can note that there is no significant image-quality degradation when considering approximate transforms.
Furthermore, in all the cases, it can be seen that~$\mathbf{\hat{C}}_4$ performs better with
a degradation of no more than~$0.52$~dB.

\begin{figure}
\centering
\includegraphics[scale=1]{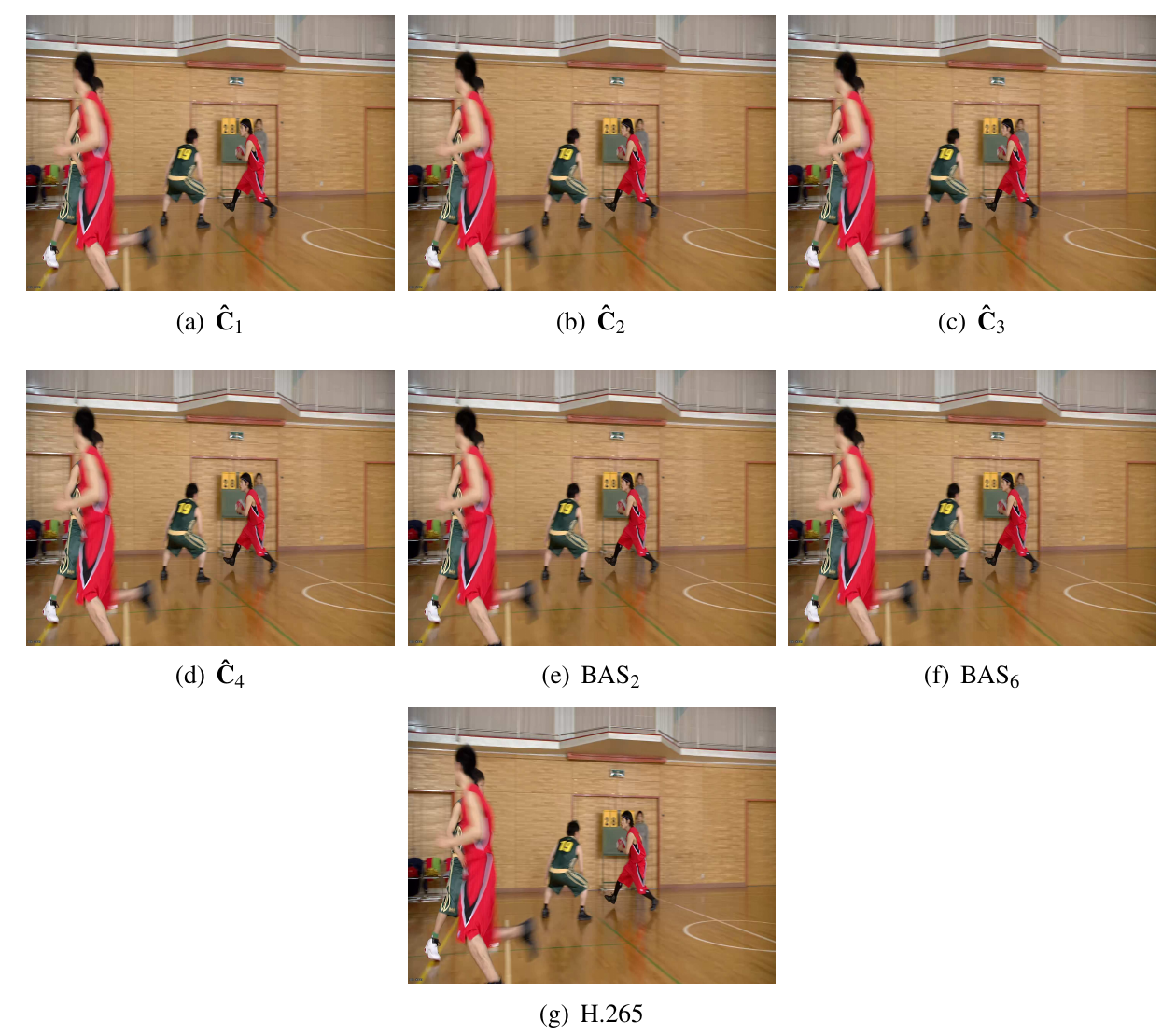}

\caption{
First frame of the H.265/HEVC-compressed `BasketballDrive' sequence with $\mbox{QP} = 32$.
}
\label{fig:exhevc}
\end{figure}

The Loeffler-based approximations could present competitive image quality while possessing low computational complexity.
Table~\ref{compcomplexHEVC} shows the arithmetic costs of both the original and modified extended versions, and the H.265/HEVC native transform.
As a qualitative result, Figure~\ref{fig:exhevc} depicts the first frame of
the `BasketballDrive' video sequence resulting from
the unmodified H.265/HEVC~\cite{hmteam} and from the modified codec according to
the discussed approximations and their scaled versions at $\text{QP}=32$.
We also display the related frames according to the BAS$_2$ and BAS$_6$ transforms.

\begin{table}

\centering
\caption{Computational complexity for
$N$-point transforms employed
in the \mbox{H.265}/ High Efficiency Video Coding (HEVC)~experiments.}
\label{compcomplexHEVC}
\begin{tabular}{c@{\:\:\:}c@{\:\:\:}c@{\:\:\:}c@{\:\:\:}c@{\:\:\:}}
\toprule
\multirow{1}{*}{\emph{\textbf{N}}} &
\multirow{1}{*}{\textbf{Transform}} &
\textbf{Add.} &
\textbf{Mult.} &
\textbf{Shifts}
\\
\midrule
\multirow{5}{*}{8}                  &  \mbox{H.265}/HEVC~\cite{Budagavi2013} &   28    &   22   &   0   \\
 &         $\mathbf{T}_1$          &   14    &   0    &  0   \\
                  &         $\mathbf{T}_3$           &   18    &   0    &  0   \\
                  &         $\mathbf{T}_5$           &   16    &   0    &  2   \\
                  &           $\mathbf{T}_6$         &   24    &   0    &  2   \\

\midrule
\multirow{5}{*}{16}                  & \mbox{H.265}/HEVC~\cite{Budagavi2013} &   100   &   86    &   0   \\
&        $\mathbf{T}_1$         &   44    &   0     &   0   \\
                  &          $\mathbf{T}_3$         &   52    &   0     &   0   \\
                  &         $\mathbf{T}_5$          &   48    &   0     &   4   \\
                  &         $\mathbf{T}_6$          &   64    &   0     &   4   \\
\midrule
\multirow{5}{*}{32}                  & \mbox{H.265}/HEVC~\cite{Budagavi2013} &   372    &   342  &  0  \\
 &      $\mathbf{T}_1$           &   120    &   0    &  0    \\
                  &         $\mathbf{T}_3$          &   136    &   0    &  0   \\
                  &     $\mathbf{T}_5$              &   128    &   0    &  8    \\
                  &     $\mathbf{T}_6$              &   160    &   0    &  8    \\
\bottomrule
\end{tabular}
\end{table}
\unskip
\begin{table}
\small
\centering
\caption{Bj\o ntegaard metrics of the modified HEVC reference software for tested video sequences.}
\label{tab:videos}
\scalebox{0.88}[0.88]{
\begin{tabular}{c@{\quad}c@{\quad}c@{\quad}c@{\quad}c@{\quad}c@{\quad}c@{\quad}c@{\quad}c@{\quad}}
\toprule
\multirow{2}{*}{\textbf{Class}\vspace{-5pt}
} & \multirow{2}{*}{\textbf{Name}\vspace{-5pt}} & \textbf{Resolution}  & \multicolumn{6}{c}{\textbf{BD-PSNR~(dB) and BD-Rate~(\%)}} \\\cmidrule{4-9}
&& \textbf{and Frame Rate} &    \boldmath{$\mathbf{\hat{C}}_1$}   &   $\mathbf{\hat{C}}_2$    &   $\mathbf{\hat{C}}_3$    &   $\mathbf{\hat{C}}_4$    &  {\textbf{BAS}}\boldmath{$_2$}    &  {\textbf{BAS}}\boldmath{$_6$}    \\
\midrule
 \multirow{2}{*}{A} & \multirow{2}{*}{’PeopleOnStreet’} & 2560 $\times$ 1600 & 0.54 dB & 0.53 dB & 0.53 dB & 0.34 dB & 0.47 dB & 0.50 dB \\
& & 30~fps & $-$9.75\% & $-$9.61\% & $-$9.49\%& $-$6.30\% & $-$8.47\% & $-$8.99\% \\\midrule%

\multirow{2}{*}{B} & \multirow{2}{*}{’BasketballDrive’} & {1920 $\times$ 1080} & 0.33 dB & 0.32 dB & 0.45 dB & 0.20 dB &0.25 dB & 0.27 dB \\

& & 50~fps & $-$11.56\% & $-$11.28\% & $-$15.52\% & $-$7.24\% & $-$8.90\% &$-$9.50\% \\\midrule%

\multirow{2}{*}{C} & \multirow{2}{*}{’RaceHorses’} & { 832 $\times$ 480} & 0.67 dB & 0.66 dB & 0.83 & 0.52 dB &0.61 dB & 0.61 dB \\
& &30~fps & $-$8.06\% & $-$7.98\% & $-$9.70\% & $-$6.36\% & $-$7.38\% & $-$7.38\% \\\midrule%

\multirow{2}{*}{D} & \multirow{2}{*}{’BlowingBubbles’} & { 416 $\times$ 240} & 0.26 dB & 0.25 dB & 0.26 dB & 0.15 dB & 0.20 dB & 0.22 dB \\
& &50~fps & $-$4.35\% & $-$4.25\% & $-$4.46\% & $-$2.59\% & $-$3.48\% & $-$3.74\% \\\midrule%

\multirow{2}{*}{E} & \multirow{2}{*}{’KristenAndSara’} & { 1280 $\times$ 720} & 0.47 dB & 0.46 dB & 0.47 dB & 0.30 dB & 0.39 dB & 0.41 dB \\

& & 60~fps & $-$8.97\% & $-$8.78\% & $-$9.00\% & $-$5.93\% & $-$7.49\% & $-$7.90\% \\\midrule%

\multirow{2}{*}{F} & \multirow{2}{*}{’BasketbalDrillText’} & { 832 $\times$ 480} & 0.20 dB & 0.20 dB &0.19 dB & 0.12 dB & 0.16 dB & 0.17 dB \\
& & 50~fps & $-$3.82\% & $-$3.79\% & $-$3.63\% & $-$2.26\% & $-$3.06\% & $-$3.28\% \\\bottomrule

\end{tabular}}
\end{table}

\section{FPGA Implementation}
\label{sec:fpga}

To compare
the hardware-resource consumption of
the discussed approximations,
they were
initially modeled and tested
in Matlab Simulink and
then physically realized on
FPGA.
The~FPGA used was a Xilinx Virtex-6 XC6VLX240T
installed on a Xilinx ML605 prototyping board.
 FPGA realization was tested with 100,000 random
eight-point input test vectors
using hardware cosimulation.
Test vectors were generated from within the Matlab environment
and routed to the physical FPGA device
using a JTAG-based hardware cosimulation.
Then, measured data from the FPGA was routed back to Matlab
memory space.

We separated the $\text{BAS}_6$ approximation
for comparison with efficient approximations
because its performance metrics
lay on the Pareto frontier
of the plots in Figure~\ref{paretolines}.
The
associated
FPGA implementations
were evaluated
for hardware complexity and real-time performance using metrics
such as
configurable logic blocks (CLB) and flip-flop (FF) count,
critical path delay ($T_\text{cpd}$) in~ns,
and maximum operating frequency ($F_\text{max}$) in~MHz.
Values were obtained from the Xilinx FPGA synthesis
and place-route tools by
accessing the \texttt{xflow.results} report file.
In addition,
 static ($Q_p$)
and
dynamic power
($D_p$ in $\mathrm{mW}/\mathrm{GHz}$)
consumption
were
estimated using the Xilinx XPower Analyzer.
We also reported
area-time complexity ($AT$)
and
area-time-squared complexity ($AT^2$).
Circuit area ($A$) was estimated using the CLB count as a metric,
and
time was derived from $T_\text{cpd}$.
Table~\ref{FPGAresults}
lists
the FPGA hardware resource and power consumption for each algorithm.
\begin{table}
\centering
\caption{Hardware-resource consumption and power consumption using a Xilinx Virtex-6 XC6VLX240T 1FFG1156 device.}
\label{FPGAresults}
\begin{tabular}{l@{\quad}c@{\quad}c@{\quad}c@{\quad}c@{\quad}c@{\quad}c@{\quad}c@{\quad}c} %
\toprule
\parbox{2cm}{\centering \textbf{Approximation}} &
\parbox{1cm}{\centering \textbf{CLB}} &
\parbox{1cm}{\centering \textbf{FF} } &
\boldmath{$T_\text{\textbf{cpd}}$ $(\mathrm{ns})$} &
\boldmath{$F_{\text{\textbf{max}}}$ $(\mathrm{MHz})$} &
\boldmath{$D_p$ $(\mathrm{mW/GHz})$} &
\boldmath{$Q_p$ $(\mathrm{W})$} &
\boldmath{$AT$} &
\boldmath{$AT^2$}\\ \midrule
 {\centering $\mathbf{T}_1$} & 108 & 368 & 2.783 & 359 & 2.93 & 3.455 & 300.56 & 836.47\\
 {\centering $\mathbf{T}_3$} & 114 & 386 & 2.805 & 357 & 3.56 & 3.462 & 319.77 & 896.95\\
 {\centering $\mathbf{T}_5$} & 125 & 442 & 2.280 & 439 & 3.65 & 3.472 & 285    & 649.8 \\
 {\centering $\mathbf{T}_6$} & 185 & 626 & 2.699 & 371 & 4.18 & 3.470 & 499.32 & 1347.65\\
{\centering $\text{BAS}_6~$\cite{Bouguezel2013}} & 116 & 316 & 2.740 & 365 & 4.05 & 3.468 & 317.84 & 870.88
\\
 \bottomrule
\end{tabular}
\end{table}

Considering the circuit complexity of
the discussed approximations and
 $\text{BAS}_6$~\cite{Bouguezel2013},
as measured from the CBL count for the FPGA synthesis report,
it can be seen from Table~\ref{FPGAresults} that $\mathbf{T}_1$ is the smallest option
in terms of circuit area.
When considering maximum speed, matrix $\mathbf{T}_5$ showed the best performance on the Vertex-6 XC6VLX240T device.
Alternatively,
if we consider the normalized dynamic power consumption,
the best performance was again measured from~$\mathbf{T}_1$.

\section{Conclusions and Final Remarks}
\label{conclusion}

In this paper,
we introduced a mathematical framework for the design of eight-point  DCT approximations.
The Loeffler algorithm
was parameterized and a class of matrices was derived.
Matrices with good properties,
such as
low-complexity,
invertibility,
orthogonality or near-orthogonality,
and
closeness to the exact DCT,
were separated
according to a multicriteria optimization problem
aiming at Pareto efficiency.
The DCT approximations in this class were assessed, and the optimal transformations were separated.

The obtained transforms were assessed
in terms of computational complexity, proximity, and coding measures.
The derived efficient solutions
constitute
DCT approximations
capable of good properties
when compared to existing DCT approximations.
At the same time,
 approximation requires extremely low computation costs:
only additions and bit-shifting operations are required
for their evaluation.

We demonstrated that the proposed method
is a unifying structure
for several approximations
scattered in the literature,
including the well-known approximation
by
Lengwehasatit--Ortega~\cite{Lengwehasatit2004},
and
they share the same
matrix factorization scheme.
Additionally,
because all discussed approximations have a common
matrix expansion,
the SFG of their associated fast algorithms
are identical
except for the parameter values.
Thus,
the resulting structure
paves the way
to performance-selectable
transforms according to the choice of parameters.

Moreover,
 approximations
were assessed and compared in terms of
image and video coding.
For images,
a JPEG-like encoding simulation
was considered,
and
for videos,
 approximations were
embedded in the \mbox{H.264}/AVC and \mbox{H.265}/HEVC video-coding standards.
Approximations exhibited good coding performance compared to exact DCT-based JPEG compression
and
the obtained frame video quality
was very close to the results shown
by the \mbox{H.264}/AVC and \mbox{H.265}/HEVC standards.

Extensions for larger blocklengths can be achieved by
considering scalable approaches such as the one suggested in Reference~\cite{Jridi2015}.
Alternatively,
one could
employ direct parameterization of the 16-point Loeffler
DCT algorithm described in Reference~\cite{Loeffler1989}.

\appendix

\section{Abbreviations}

The following abbreviations are used in this manuscript:

\noindent
\begin{tabular}{@{}ll}
DCT & Discrete cosine transform\\
AVC & Advanced video coding\\
HEVC & High-efficiency video coding\\
FPGA & Field-programmable gate array\\
KLT & Karhunen--Lo\`eve transform \\
%DWT & Discrete wavelet transform\\
%JPEG & Joint photographic experts group\\
%MPEG & Moving picture experts group\\
%MSE & Mean-squared error\\
%SSIM & Structural similarity index metric\\
%PSNR & Peak signal-to-noise ratio\\
%WHT & Walsh-Hadamard transform \\
%QP & Quantization parameter\\
%BD &  Bj{\o}ntegaard delta\\
%CIF & Common intermediate format\\
%CLB & Configurable logic block\\
%FF & Flip-flop
%DCT & Discrete cosine transform\\
%AVC & Advanced video coding\\
%HEVC & High efficincy video coding\\
%%
%%
%FPGA & Field programmable gate array\\
%KLT & Karhunen-Lo\`eve transform \\
DWT & Discrete wavelet transform\\
JPEG & Joint photographic experts group\\
MPEG & Moving-picture experts group\\
MSE & Mean-squared error\\
SSIM & Structural similarity index metric\\
PSNR & Peak signal-to-noise ratio\\
WHT & Walsh--Hadamard transform \\
QP & Quantization parameter\\
BD &  Bj{\o}ntegaard delta\\
CIF & Common intermediate format\\
CLB & Configurable logic block\\
FF & Flip-flop
\end{tabular}

{\small
\singlespacing
\bibliographystyle{siam}
\bibliography{bibcleanoutput}
}

\end{document}